\documentclass[a4paper,fleqn,usenatbib]{mnras}

\usepackage[T1]{fontenc}
\usepackage{ae,aecompl}

\usepackage{graphicx}	
\usepackage{amsmath}	
\usepackage{amssymb}	
\usepackage{multicol}
\usepackage{bm}
\usepackage{pdflscape}
\usepackage{relsize}
\usepackage[normalem]{ulem}

\title[$H$-band Temperature and Metallicity indicators]{\textbf{$H$-band Temperature and Metallicity Indicators for Cool Giants \\ ~~~~~~~~~~~~~~~~~~~\smaller{Empirical Relations in Bayesian Framework}}}
\author[Ghosh et al. 2019]{
Supriyo Ghosh,$^{1}$\thanks{E-mail: supriyo.ghosh@tifr.res.in (SG)},
J. P. Ninan$^{2,3}$, and D. K. Ojha$^{1}$
\\
$^{1}$Tata Institute of Fundamental Research, Homi Bhabha Road, Colaba, Mumbai-400 005, India \\
$^{2}$Department of Astronomy \& Astrophysics, 525 Davey Laboratory, The Pennsylvania State University, University Park, PA 16802, USA\\
$^{3}$ Center for Exoplanets and Habitable Worlds, 525 Davey Laboratory, The Pennsylvania State University,University Park, PA 16802, USA\\
}

\date{Accepted XXX. Received YYY; in original form ZZZ}

\pubyear{2021}

\begin{document}
\label{firstpage}
\pagerange{\pageref{firstpage}--\pageref{lastpage}}
\maketitle

\begin{abstract}
We explored here the near-infrared $H$-band atmospheric window aiming to provide quantitative diagnostic tools for deriving stellar parameters, for instance, effective temperature ($T_{eff}$) and metallicity ([$Fe/H$]), of cool giants ($T_{eff}$ $<$ 5000 K) using low-resolution spectra. We obtained 177 cool giants from the X-shooter spectral library covering a wider metallicity range ($-$2.35 dex $<$ [$Fe/H$] $<$ 0.5 dex) than in earlier works. Degrading the spectral resolution to R$\sim$ 1200, we estimated equivalent widths of several important spectral features, and the behavior of spectral features with stellar parameters are studied. Also, the empirical relations for deriving $T_{eff}$ and [$Fe/H$] are established in the Bayesian framework. We found that $^{12}$CO at 1.56 $\mu$m and 1.62 $\mu$m, and $^{12}$CO+Mg{\small I} at 1.71 $\mu$m are the best three $T_{eff}$ indicators with a typical accuracy of 153 K, 123 K and 107 K, respectively. The cubic Bayesian model provides the best metallicity estimator with a typical accuracy of 0.22 dex, 0.28 dex, and 0.24 dex for FeH at 1.62 $\mu$m, $^{12}$CO at 1.64 $\mu$m, and  Fe {\small I} at 1.66 $\mu$m, respectively. We also showed a detailed quantitative metallicity dependence of $T_{eff}$--EWs correlations defining three metallicity groups, supersolar ([$Fe/H$] $>$ 0.0 dex), solar ($-$0.3 dex $<$ [$Fe/H$] $<$ 0.3 dex), and subsolar ([$Fe/H$] $<$ $-$0.3 dex), from Hierarchical Bayesian modelling. The difference between the solar and subsolar relationship is statistically significant, but such difference is not evident between the solar and supersolar groups.
\end{abstract}




\section{Introduction}
	
Precise and accurate estimation of stellar parameters (e.g., $T_{eff}$, log $g$ and [$Fe/H$]) for cool giants is important to characterize stellar populations in different Galactic and extragalactic environments. However, such estimation is a long-standing challenge because of the complex molecular near-photospheric environment of cool giants. The Near-Infrared (NIR) spectral region is of much interest for studying cool giants ($T_{eff}$ $\sim$ 3000--5000 K) as they emit maximum energy (peak near 1 $\mu$m) in the NIR. Also, the NIR regime suffers relatively less extinction than the optical allowing us to probe long distances in the Galaxy. The majority of previous works especially focus on the NIR $K$-band spectral features such as Na I at 2.20 $\mu$m, Ca I at 2.26 $\mu$m and $^{12}$CO at 2.29 $\mu$m sensitive to $T_{eff}$, log $g$ and [$Fe/H$] for estimating stellar parameters \citep{1986apjs..62..501, 1993aap..280..536, 1997apjs..111..445, 1997AJ..113..1411, 1998ApJ...508..397M, 2000aj..120..2089, 2000aaps..146..217, 2000AJ....120..833, 2001AJ..122..1896, 2004apjs..151..387, 2006A&A...458..609D, 2011apj..741..108, 2013aap...549A.129, 2016aap..590A..6, 2019MNRAS.484.4619G,2021MNRAS.501.4596G}. However, at $\lambda$ $>$ 1.8 $\mu$m, the presence of hot dust and active galactic nuclei (AGN) contribute significantly towards the continuum emission and can heavily veil stellar photospheres \citep{1998ApJ...508..397M}. Thus, $K$-band spectral features could be diluted and become unsuitable for estimating stellar parameters. On the other hand, such contamination of spectral features becomes insignificant at $\lambda$ $<$ 1.8 $\mu$m making the $H$-band atmospheric window suitable for stellar population studies. In addition, the minimum opacity at 1.6 $\mu$m allows us to probe deepest into the stellar photosphere (e.g., \citealt{Lancon1992}).  

Since the pioneering work of \citet{1970aj..75..785}, a limited number of works have been done to explore the behaviour of spectral features in the $H$-band atmospheric window with stellar parameters at different spectral resolutions (e.g. \citealt{Lancon1992, 1993aap..280..536, 1996A&AS..116..239D, 1998ApJ...508..397M, 2004apjs..151..387, 2009apjs..185..289, 2017apjs..230..23, 2018ApJS..238...29P, 2019MNRAS.484.4619G}). These studies reveal that spectra in the $H$-band atmospheric window are very complex for cool giants and thus, the identification of spectral features is a challenge. This is because of the presence of many relatively weak metal absorption features (e.g., Mg I at 1.50 $\mu$m and 1.71 $\mu$m, Fe I at 1.60 $\mu$m and 1.65 $\mu$m, and Al I at 1.67 $\mu$m), the second CO overtone vibrational-rotational bands, and OH lines. Some of these spectral features are found to be good indicators for $T_{eff}$ (e.g., $^{12}$CO at 1.62 $\mu$m, Al I at 1.67 $\mu$m), log $g$ ($^{12}$CO at 1.62 $\mu$m) and luminosity class (e.g., ratio of $^{12}$CO at 1.62 $\mu$m to Si I at 1.59 $\mu$m, ratio of $^{12}$CO at 1.62 $\mu$m to sum of Mg I at 1.50 $\mu$m and 1.71 $\mu$m) and thus, these features can be used for stellar population studies \citep{1993aap..280..536, 1998ApJ...508..397M, 2004apjs..151..387, 2018ApJS..238...29P, 2019MNRAS.484.4619G}. \citet{1993aap..280..536} was the first to study the comparative behaviour of $H$-band indices (Si I at 1.59 $\mu$m and CO at 1.62 $\mu$m) with $K$-band CO at 2.29 $\mu$m and demonstrated the advantages of using spectral indices at around 1.6 $\mu$m in presence of hot dust from the measurements of the Seyfert galaxy NGC 1068. Subsequently, a few empirical relations were established between various spectral features and stellar parameters -- $T_{eff}$ and log $g$ (see, \citealt{1998ApJ...508..397M, 2004apjs..151..387, 2018ApJS..238...29P, 2019MNRAS.484.4619G}). A very recent study of \citet{2020A&A...641A..44M} deserves special mention who presented a qualitative analysis of a number of spectral indices with stellar parameters for $H$-band atmospheric window as well as for $Y$, $J$, and $L$-bands atmospheric windows using the low-resolution (R $\sim$ 2000) spectra from Infrared Telescope Facility (IRTF) spectral library \citep{2009apjs..185..289}. In addition, the $H$-band atmospheric window has been extensively acknowledged in literature to derive chemical abundances of cool giants at different Galactic locations because of the rich concentration of both metal and molecular lines (e.g., \citealt{2004AJ....127.3422O, 2010A&A...509A..20R, 2015A&A...574A..80V,2020MNRAS.491..544B}). The ongoing large-scale high-resolution (R $\sim$ 22500) stellar spectroscopic survey like Apache Point Observatory Galactic Evolution Experiment (APOGEE) is looking for all Milky Way components for detailed chemical and kinematic information using the $H$-band atmospheric window (1.51--1.70 $\mu$m) of the electromagnetic spectrum. To summarize, we opine that although previous studies of $H$-band atmospheric window are more limited than those in the $K$-band window, the former offers many important diagnostic spectral features for characterizing different stellar populations. However, accurate, prior knowledge of the behaviour of the spectral features with a broad parameter coverage, precisely metallicity, is required for the precise characterization.

\begin{figure}
	\centering
	\includegraphics[scale=0.55]{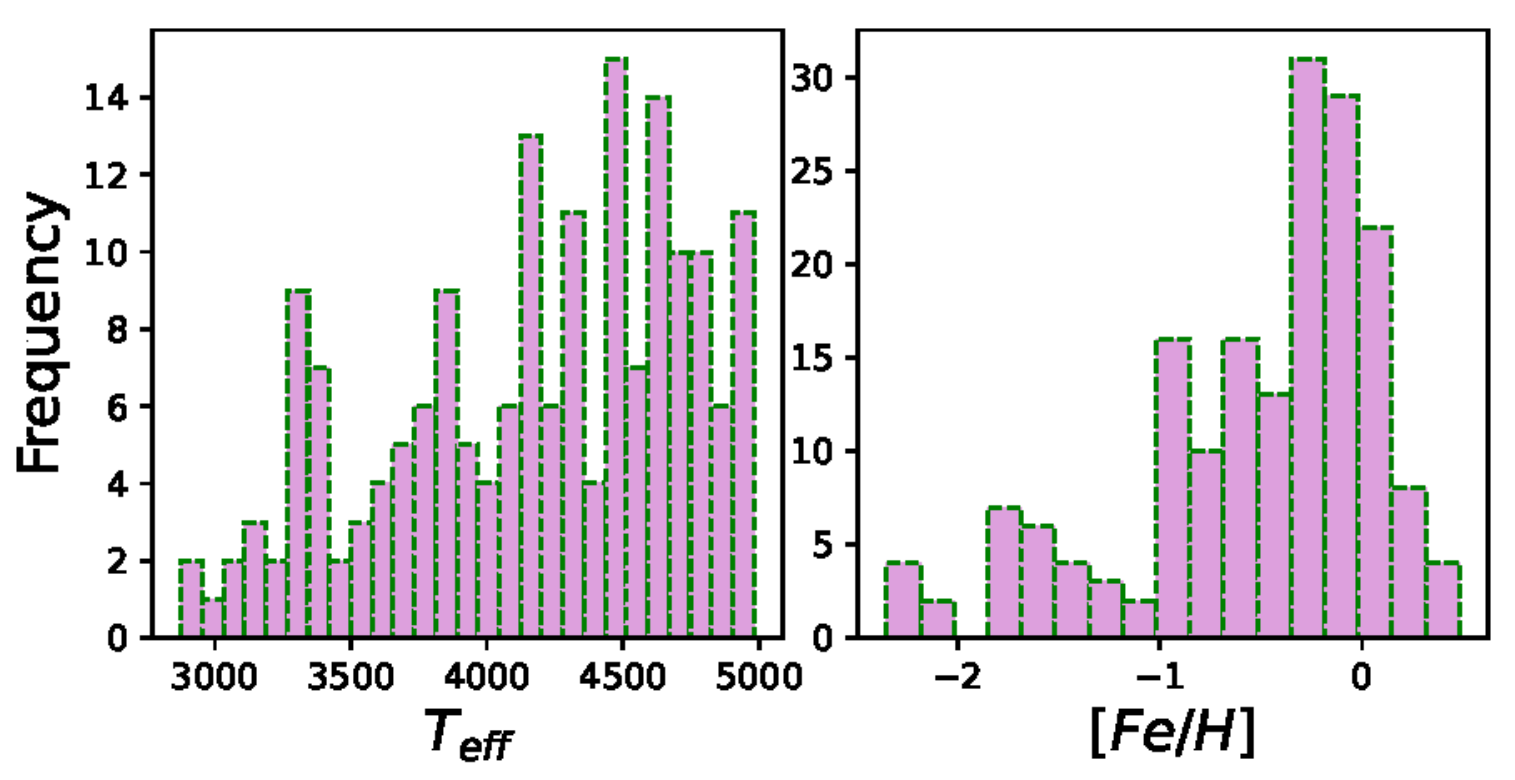}
	\caption{Histograms of the stellar parameters (Left-hand panel: $T_{eff}$, and right-hand panel: [$Fe/H$]) for 177 cool giants.}
	\label{Fig:sample_selection}
\end{figure}

\begin{table*}
	\centering
	\caption{Stellar parameters and estimated EWs of the spectral features.}
	\label{tab:Stellar_parameters_and_estimated_EWs}
	
	\resizebox{0.99\textwidth}{!}{
		\begin{tabular}{lccccccccccc} 
			\hline
			\hline
			Stars names & $T_{eff}$  & log $g$  & [$Fe/H$] & EW$_{CO156}$ & EW$_{Si159}$  & EW$_{CO162}$ & EW$_{FeH162}$ & EW$_{CO164}$ & EW$_{Fe166}$ & EW$_{Al167}$ & EW$_{COMg171}$ \\
			\hline
			
			ISO-MCMS J004950.3$-$731116 & 3827 $\pm$ 52 & 0.68 $\pm$ 0.19 & $-$0.52 $\pm$ 0.17 & 3.955 $\pm$ 0.428 & 3.738 $\pm$ 1.128 & 3.777 $\pm$ 0.463 & 1.667 $\pm$ 0.63 & 4.009 $\pm$ 0.85 & 0.685 $\pm$ 0.465 & 3.196 $\pm$ 0.88 & 5.072 $\pm$ 0.681 \\
			ISO-MCMS J005059.4$-$731914 & 3806 $\pm$ 51 & 0.71 $\pm$ 0.19 & $-$0.92 $\pm$ 0.17 & 5.709 $\pm$ 0.789 & 3.76 $\pm$ 1.253 & 4.887 $\pm$ 0.763 & 2.195 $\pm$ 0.282 & 4.83 $\pm$ 1.626 & 0.31 $\pm$ 0.636 & 3.036 $\pm$ 1.037 & 3.718 $\pm$ 0.647 \\
			$[$M2002$]$ SMC 83593 & 3607 $\pm$ 59 & 0.88 $\pm$ 0.19 & $-$0.98 $\pm$ 0.17 & 4.783 $\pm$ 0.764 & 4.288 $\pm$ 0.685 & 3.96 $\pm$ 0.876 & 1.621 $\pm$ 0.575 & 3.737 $\pm$ 0.628 & 0.869 $\pm$ 0.564 & 3.276 $\pm$ 0.712 & 5.515 $\pm$ 1.446 \\
			ISO-MCMS J005314.8$-$730601 & 3762 $\pm$ 38 & 1.12 $\pm$ 0.14 & $-$0.71 $\pm$ 0.09 & 4.227 $\pm$ 0.431 & 4.001 $\pm$ 0.879 & 3.655 $\pm$ 0.566 & 1.498 $\pm$ 0.404 & 2.923 $\pm$ 0.88 & 0.574 $\pm$ 0.391 & 3.183 $\pm$ 1.159 & 5.979 $\pm$ 0.486 \\
			ISO-MCMS J005332.4$-$730501 & 4391 $\pm$ 32 & 1.05 $\pm$ 0.14 & $-$0.58 $\pm$ 0.06 & 1.21 $\pm$ 0.453 & 3.275 $\pm$ 0.277 & 1.284 $\pm$ 0.224 & $-$0.375 $\pm$ 0.169 & 2.035 $\pm$ 0.366 & 0.548 $\pm$ 0.323 & 1.327 $\pm$ 0.486 & 2.453 $\pm$ 0.793 \\
			... & ... & ... & ... & ... & ... & ... & ... & ... & ... & ... & ...  \\
			\hline
			
	\end{tabular}} \\
	\textit{Notes}: Table \ref{tab:Stellar_parameters_and_estimated_EWs} originally contains EWs (in \AA) of all spectral features estimated in this work along with stellar parameters of the sample. Table \ref{tab:Stellar_parameters_and_estimated_EWs} is available in its entirety in the electronic version of the journal as supplementary material. However, only a portion of the table is shown here to show its' form and content.	
\end{table*} 

\begin{table*}
	\centering
	\caption{Spectral bands for EWs estimation.}
	\label{tab:bandpass}
	\resizebox{0.9\textwidth}{!}{
		\begin{tabular}{crccc} 
			\hline
			\hline
			Index & Feature & Feature  & Continuum  &  References$^a$ \\	 
			&         & Bandpass ($\mu$m) & Bandpass ($\mu$m) &  \\
			\hline		
			
			Mg150 & Mg I at 1.50 $\mu$m & 1.5000--1.5080 & 1.4920--1.5080, 1.5100--1.5120 & 1, 2, 3 \\
			K152 & K I at 1.52 $\mu$m & 1.5150--1.5200 & 1.5100--1.5120, 1.5230--1.5250 & 3 \\
			Fe153 & Fe I  at 1.53 $\mu$m & 1.5280--1.5330 & 1.5230--1.5250, 1.5480--1.5500 & 3 \\
			H155 & H I at 1.55 $\mu$m & 1.5520--1.5570 & 1.5480--1.5500, 1.5930--1.5940 & 3\\
			CO156 & $^{12}$CO at 1.56 $\mu$m & 1.5570--1.5635 & 1.5480--1.5500, 1.5930--1.5940 & 3 \\
			H157 & H I at 1.57 $\mu$m & 1.5670--1.5720 & 1.5480--1.5500, 1.5930--1.5940 & 3 \\
			Mg157 & Mg I at 1.57 $\mu$m & 1.5730--1.5800 & 1.5480--1.5500, 1.5930--1.5940 & 2, 3, 4\\
			FeH158 & FeH at 1.58 $\mu$m & 1.5820--1.5860 & 1.5480--1.5500, 1.5930--1.5940 & 1, 2, 3, 5\\
			Si159 & Si I at 1.59 $\mu$m & 1.5870--1.5910 & 1.5480--1.5500, 1.5930--1.5940 & 1, 2, 3, 4, 5, 6\\
			CO160 & $^{12}$CO at 1.60 $\mu$m & 1.5950--1.6000 & 1.5930--1.5940, 1.6160--1.6180 & 2, 3\\
			Fe161 & Fe I at 1.61 $\mu$m & 1.6050--1.6090 & 1.5930--1.5940, 1.6160--1.6180 & 3 \\
			H161 & H I at 1.61 $\mu$m & 1.6100--1.6140 & 1.5930--1.5940, 1.6160--1.6180 & 3\\
			CO162 & $^{12}$CO at 1.62 $\mu$m & 1.6180--1.6220 & 1.6160--1.6180, 1.6340--1.6370 & 3, 4, 5, 6 \\
			FeH162 & FeH at 1.62 $\mu$m & 1.6240--1.6280 & 1.6160--1.6180, 1.6340--1.6370 & 3 \\
			CO164 & $^{12}$CO 1.64 $\mu$m & 1.6390--1.6470 & 1.6340--1.6370, 1.6585--1.6605 & 3\\
			Fe166 & Fe I at 1.66 $\mu$m & 1.6510--1.6580 & 1.6340--1.6370, 1.6585--1.6605 & 3\\
			CO166 & $^{12}$CO at 1.66 $\mu$m & 1.6605--1.6640 & 1.6585--1.6605, 1.6775--1.6790 & 3, 4\\
			Al167 & Al I at 1.67 $\mu$m & 1.6705--1.6775 & 1.6585--1.6605, 1.6775--1.6790 & 3, 4\\
			H168 & H I at 1.68 $\mu$m & 1.6790--1.6825 & 1.6775--1.6790, 1.6825--1.6835 & 3, 4\\
			COMg171 & $^{12}$CO+MgI at 1.71 $\mu$m & 1.7050--1.7130 & 1.6920--1.6960, 1.7140--1.7160 & 2, 3, 4 \\
			\hline
		\end{tabular}
	} 
	\\
	\textit{References}: (1) \citet{2004apjs..151..387}; (2) \citet{2019MNRAS.486.3228R}; (3) \citet{2020A&A...641A..44M}; (4) \citet{1998ApJ...508..397M}; \\ (5) \citet{1993aap..280..536}; (6) \citet{2019MNRAS.484.4619G}. \\
	\textit{Note}: $^a$References: Literature that studied the similar index.
\end{table*}

\begin{figure*}
	\centering
	\includegraphics[scale=0.72]{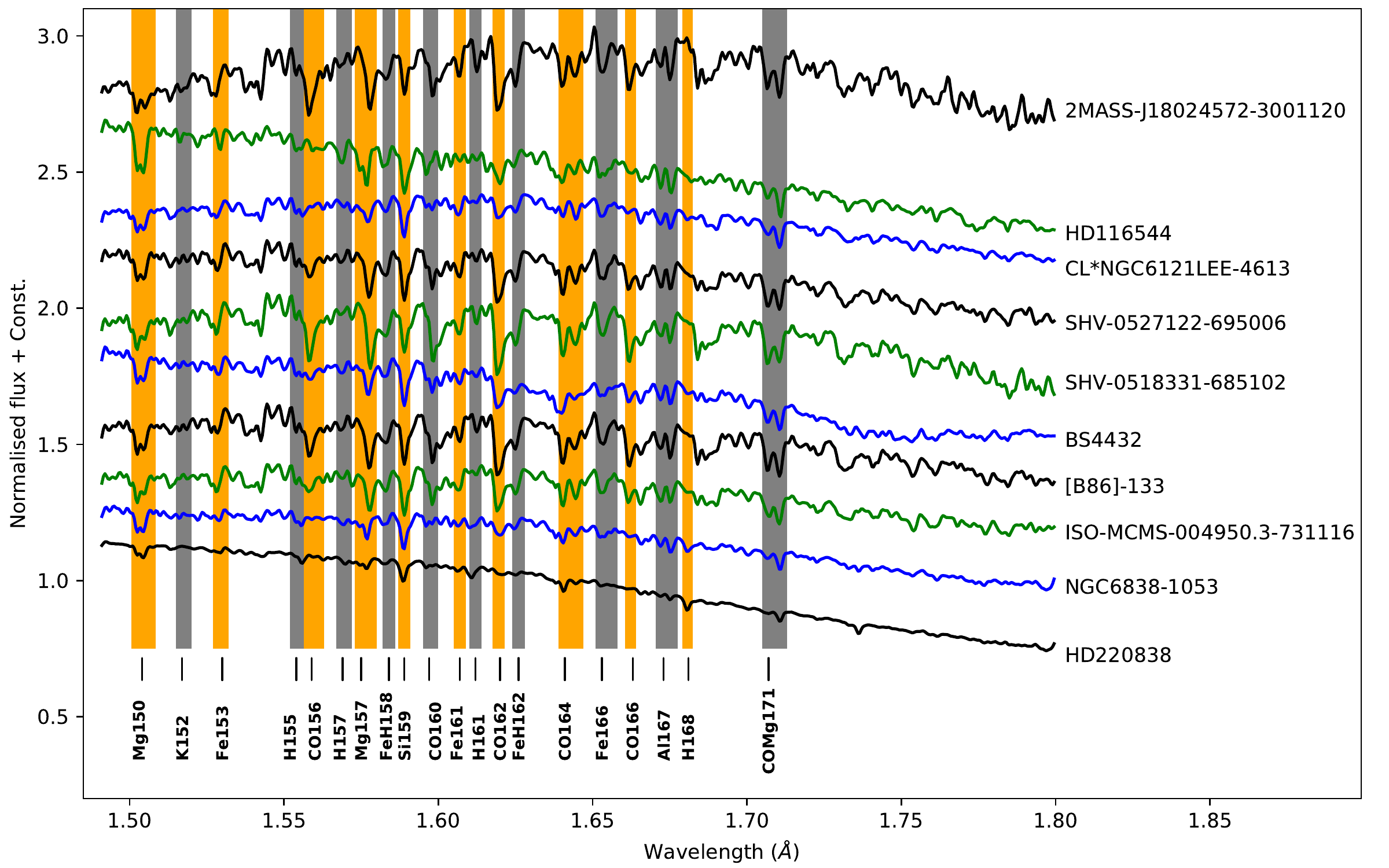}
	\caption{Figure shows a subset of the cool giants spectra used for this study, obtained by reducing the resolution to R $\sim$ 1200 from publicly available X-shooter spectral library.			
			All the spectra have been normalized
			to unity at 1.65 $\mu$m and offset by constant value with respect to the bottom-most spectrum for displaying purposes. The names of the stars have been mentioned at the right end of the corresponding spectra. All the prominent features are marked. The coloured regions represent the feature bandpasses as listed in Table \ref{tab:bandpass}.}
	\label{Fig:sample_spectra}
\end{figure*}

In this work, we aim to provide easy-to-use reliable diagnostic tools based on the $H$-band spectral features of well-characterized sample stars. The beauty of such diagnostic tools is that they are independent of any knowledge regarding the reddening and distance to the object. Also, we can derive stellar parameters very easily by measuring the line strength of spectral features. While we do not expect the precision to be as good as modelling the high resolution spectra, we expect these relations to be useful for estimating star properties, as well as confidence intervals on those estimates, for giant stars' data from large low-resolution spectroscopic surveys. These empirical relationships will also be useful for stars which are not bright enough for high-resolution spectroscopy. Earlier works established empirical relations between spectral features and stellar parameters as stated above, however, the key limitation of earlier spectral libraries is a small number of stars with a narrow metallicity range. Thus, the behaviour of spectral features with metallicity was hardly explored. In this context, the second data release of the X-shooter stellar library \citep{2020A&A...634A.133G} provides a wide metallicity coverage (--2.35 dex $<$ [$Fe/H$] $<$ 0.5 dex, \citealt{2019A&A...627A.138A}). The library, thus, enables to explore possible empirical relations between spectral features and parameters for a wide range of metallicity and study possible metallicity dependence on those empirical relations. Besides, this work helps to understand how precisely stellar parameters such as $T_{eff}$ and [$Fe/H$] can be obtained from low-resolution (R $\sim$ 1200) $H$-band spectra. This would be highly valuable to understand the usefulness of low-resolution spectrographs for evaluating giant star properties. The paper is organized as follows. Section~\ref{Data_collection} describes the sample giants used for this work and section~\ref{Result_and_Discussion} deals with our new results and discussion. Finally, the summary of the work and conclusions are drawn in section~\ref{Summary_and_Conclusions}.   

\section{Sample selection} \label{Data_collection}
We made use of the second data release of the X-shooter stellar library \citep{2020A&A...634A.133G}, which contains 813 spectra for 666 stars of various spectral types. The spectra were observed in the wavelength range 0.3--2.5 $\mu$m at a spectral resolution ($R$) $\sim$ 10000 using the X-shooter spectrograph on ESO’s VLT  \citep{2011A&A...536A.105V}. These stars are located at different Galactic locations such as in star clusters, Galactic disc, in the Galactic bulge, and in the Magellanic Clouds. The details about sample selection, observing strategy, data reduction, and calibration can be found in \citet{2020A&A...634A.133G}. The library contains a total of 381 cool giants ($T_{eff}$ $<$ 5000 K). However, a number of stars (141 stars) were retrieved as supergiants, Mira variables, and OH/IR stars from the SIMBAD database. These stars are rejected from our sample to minimize the selection bias as supergiants have stronger bands strength than normal giants and spectra of long-period variables (Mira and OH/IR stars) show pulsation dependent variation \citep{2000A&AS..146..217L,2018aj..155..216,2021AJ....161..198G}. Furthermore, 33 stars out of the remaining 240 are observed more than once. Thus, a total of 177 cool giants were obtained for this work. We adopted $T_{eff}$ and [$Fe/H$] of sample giants from \citet{2019A&A...627A.138A} that are derived using the full-spectrum fitting package University of Lyon Spectroscopic analysis Software (ULYSS, \citealt{2009A&A...501.1269K}) and the Medium-resolution INT Library of Empirical Spectra (MILES) library \citep{2006MNRAS.371..703S, 2011A&A...532A..95F}. We refer to \citet{2019A&A...627A.138A} for additional details about the parameter-estimation procedure. Stars with more than one observation, the straight mean of the various measurements is taken. The uncertainties in estimation are 26--132 K and 0.14--0.21 dex in $T_{eff}$ and [$Fe/H$], respectively. Fig.~\ref{Fig:sample_selection} displays $T_{eff}$ and [$Fe/H$] distribution of the sample, and the parameters of the sample stars are listed in Table \ref{tab:Stellar_parameters_and_estimated_EWs}. Subsets of sample giants are presented in Fig.~\ref{Fig:sample_spectra}. Our sample spans in a wider range of stellar parameters space ($T_{eff}$ $\sim$ 3000--5000 K and [$Fe/H$] $\sim$ $-$2.35 to +0.5 dex) than in the past. Because of the wide range of parameter distribution, we can establish empirical relations between spectral features and stellar parameters for a wide range of metallicity and explore possible metallicity dependence on those empirical relations.

In this work, we selected the $H$-band atmospheric window and computed equivalent widths (EWs) of a number of prominent spectral features, Mg I at 1.50 $\mu$m (Mg150), K I at 1.52 $\mu$m (K152), Fe I  at 1.53 $\mu$m (Fe153), H I at 1.55 $\mu$m (H155), $^{12}$CO at 1.56 $\mu$m (CO156), H I at 1.57 $\mu$m (H157), Mg I at 1.57 $\mu$m (Mg157), FeH at 1.58 $\mu$m (FeH158), Si I at 1.59 $\mu$m (Si159), $^{12}$CO at 1.60 $\mu$m (CO160), Fe I at 1.61 $\mu$m (Fe161), H I at 1.61 $\mu$m (H161), $^{12}$CO at 1.62 $\mu$m (CO162), FeH at 1.62 $\mu$m (FeH162), $^{12}$CO at 1.64 $\mu$m (CO164), Fe I at 1.66 $\mu$m (Fe166), $^{12}$CO at 1.66 $\mu$m (CO166), Al I at 1.67 $\mu$m (Al167), H I at 1.68 $\mu$m (H168), and $^{12}$CO + Mg I at 1.71 $\mu$m (COMg171), to measure the line strength. These features are marked in Fig.~\ref{Fig:sample_spectra}. To estimate EWs, \\
(i) feature and continuum bands are adopted from various literature as listed in Table~\ref{tab:bandpass}. All spectral features of our interest and their corresponding bandpasses are alike to the study of \citet{2020A&A...641A..44M}. However, our work represents a step forward with respect to the works by \citet{2020A&A...641A..44M}, which was based on the small number of cool giants with a narrow metallicity range (from $-$0.6 to 0.4 dex) of IRTF spectral library. \\ 
(ii) We degraded spectral resolution from $R$ $\sim$ 10000 to $R$ $\sim$ 1200 before EWs measurement as our motivation in this work is to evaluate how accurate stellar parameters can be derived from fairly low-resolution spectra. \\
(iii) The spectral features were corrected for the zero velocity by shifting.\\ 
Finally, EWs and their uncertainties are estimated following the method as described by \citet{2014AJ....147...20N}. These estimated EWs are used to study the behavior of spectral features with stellar parameters in the following section. The estimated EWs and their uncertainties are listed in Table \ref{tab:Stellar_parameters_and_estimated_EWs}.

\section{Results and Discussion} \label{Result_and_Discussion}
 
\begin{figure*}
	\centering
	\includegraphics[width=17cm,height=10.58cm]{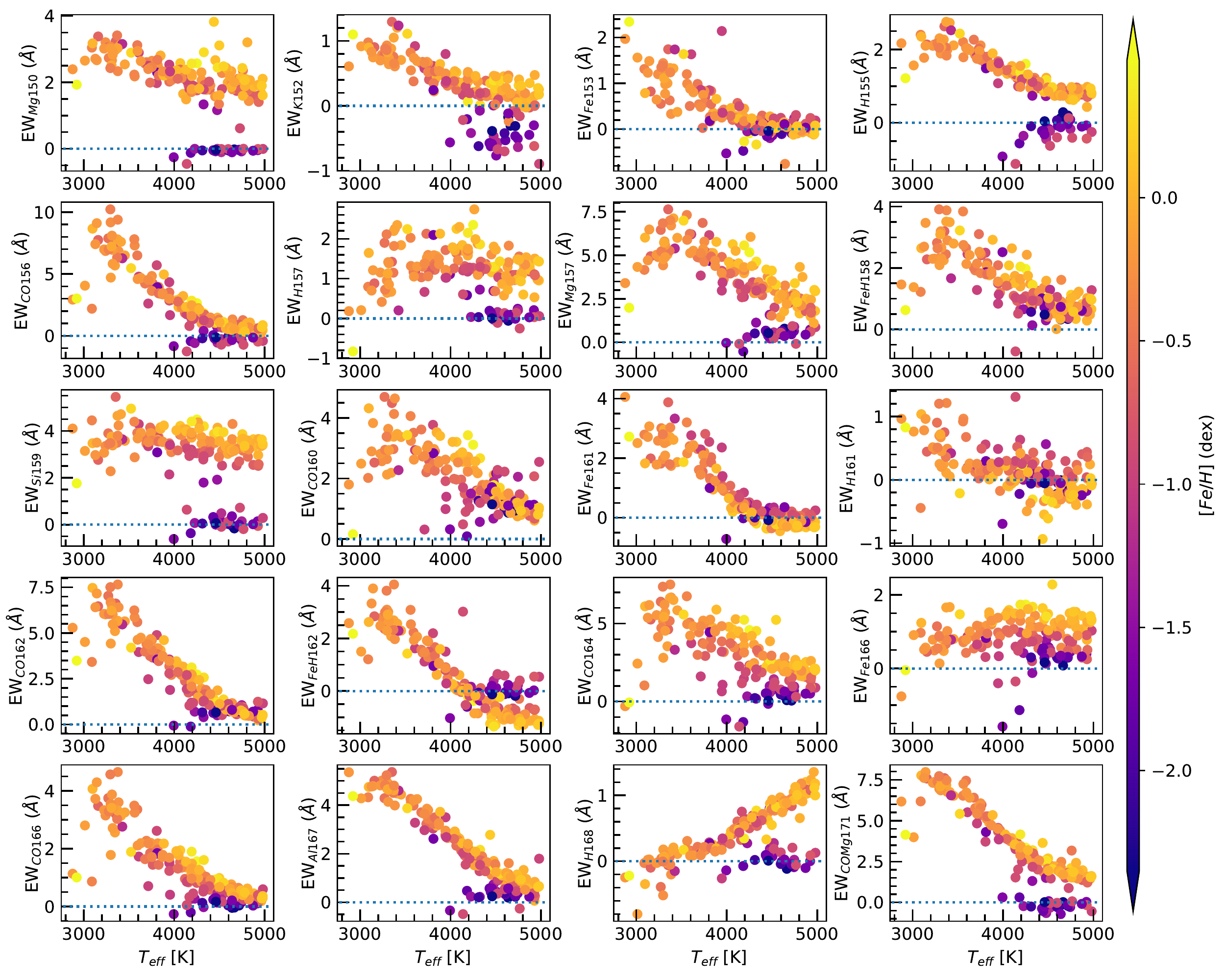}
    \caption{Variations of the $H$-band spectral features as a function of effective temperature. See the text for the detailed discussion. The colour bar represents the metallicity of each star. The dotted line separates positive and negative EWs on the plot.}
     \label{Fig:index-Teff_behaviour}
\end{figure*}

\begin{figure*}
	\includegraphics[width=17cm,height=10.6cm]{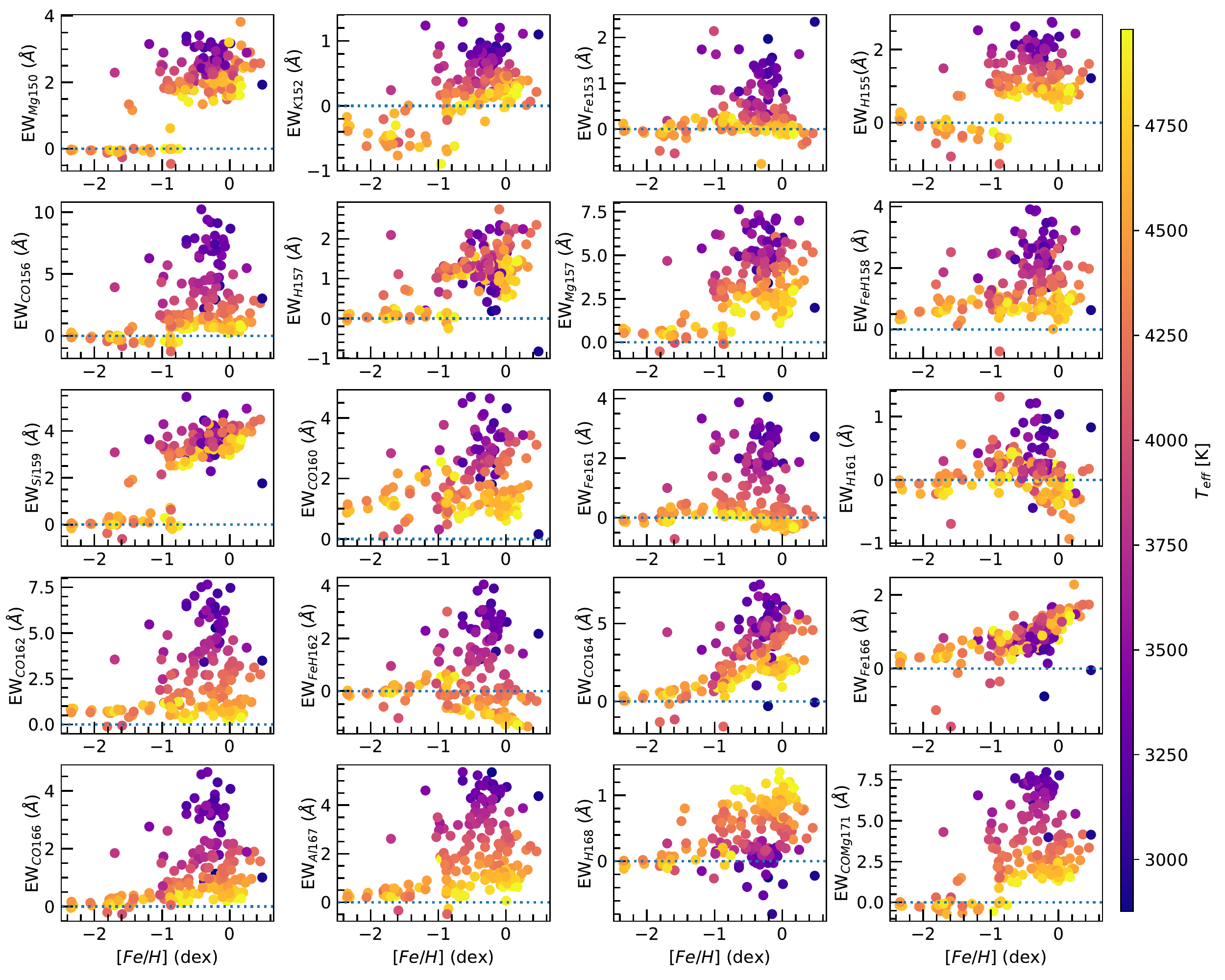}
    \caption{Variations of the $H$-band spectral features as a function of metallicity discussed in the text. The colour bar represents the effective temperature of each star. The dotted line separates positive and negative EWs on the plot.}
    \label{Fig:index-meta_behaviour}
\end{figure*}

\subsection{Behaviour of selected features with stellar parameters}
We first studied the behaviour of all relevant spectral features as mentioned in the earlier section with stellar parameters ($T_{eff}$ and [$Fe/H$]). The behaviour of those lines with $T_{eff}$ and [$Fe/H$] is presented in Figs. \ref{Fig:index-Teff_behaviour} and \ref{Fig:index-meta_behaviour}. It can be seen that the measured EWs have negative to positive values. EWs of zero or negative values represent very weak or absence of specific lines. We did not modify negative EWs to zero to preserve the statistical nature of the data. Only measurements with positive EWs were considered for further study. These data points lie above the dotted line as shown in Figs. \ref{Fig:index-Teff_behaviour} and \ref{Fig:index-meta_behaviour}.

As can be seen from Fig.~\ref{Fig:index-Teff_behaviour}, the majority of indices (Mg150, K152, Fe153, H155, CO156, Mg157, FeH158, CO160, Fe161, CO162, FeH162, CO164, CO166, Al167, H168 and COMg171) show a strong dependence on $T_{eff}$, while  H157, Si159, H161 and Fe166 show weak or no dependence on $T_{eff}$. The strong dependence of spectral lines on $T_{eff}$ is however over different temperature and metallicity ranges, for example, Mg150, K152, H155, Mg157, COMg171 at [Fe/H] $<$ -1.5 dex, Fe153, Fe161 and FeH152 at $T_{eff}$ < 4500 K and CO156, FeH158, CO160, CO162, CO164, CO166, Al167 over the entire parameter space of the sample. Furthermore, Mg150, K152, Fe153, H155, CO156, Mg157, FeH158, CO160, Fe161, CO162, FeH162, CO164, CO166, Al167, and COMg171 lines display a negative correlation with $T_{eff}$ i.e. the line strength increases as $T_{eff}$ decreases, while H168 shows a positive correlation.  However, some of them display a large scatter. This may be because of the large metallicity coverage of our sample stars. Eyeballing $T_{eff}$--EWs behaviour presented in Fig.~\ref{Fig:index-Teff_behaviour} shows very tight correlation for K152, H155, CO156, CO162, FeH162, CO166, Al167, H168 and COMg171. These lines can be used as a potential $T_{eff}$ indicator. In fact, previous studies such as \citet{1993aap..280..536, 1998ApJ...508..397M, 2004apjs..151..387, 2018ApJS..238...29P, 2019MNRAS.484.4619G} already found that some of the lines such as CO156, CO162, Al167, and COMg171 are very sensitive to $T_{eff}$. Here, we confirmed the same trends for those lines with sample stars having a large coverage in the metallicity space. Furthermore, CO156 appears to be the strongest feature in the $H$-band atmospheric window in this study, however, it is unlikely to be the case as described by \citet{1993aap..280..536} (see Sec 3.1 of the paper). The possible reason of such discrepancy could be the significant contribution coming from Fe {\small I} at 1.56 $\mu$m line.   
  
 A variation of the EWs with metallicity is also evident in Fig.~\ref{Fig:index-meta_behaviour}. A strong correlation is noticeable for FeH162, CO164, and Fe166 over the almost entire range of metallicity. Thus, these features can be used as a possible metallicity indicator in the $H$-band atmospheric window. However, a few giants with [$Fe/H$] $\sim$ 0.0 dex do not follow the same metallicity trends for FeH162 and CO164, rather they appear as a bimodal-like distribution. We found that these stars are mainly relatively warmer giants ($T_{eff}$ $\geq$ 4500 K) with log $g$ $>$ 2.0 dex. Thus, the line strength of these stars becomes weaker in comparison to other giants at the same metallicity. Also, Si159 shows monotonic trends on metallicity below [$Fe/H$] $<$ $-$1.2 dex. However, Si159 is severely contaminated with Fe I line at 1.60 $\mu$m \citep{2020A&A...641A..44M}. Furthermore, other spectral indices investigated here have weak or no correlation with metallicity. Even the metallic lines such as Mg150, Mg157, Fe161, Al167, and COMg171 rather serve as $T_{eff}$ indicators than [$Fe/H$]. This may be due to contamination by other lines.
 
 \subsection{Empirical relations}
\begin{figure}
	\centering
	\includegraphics[scale=0.3]{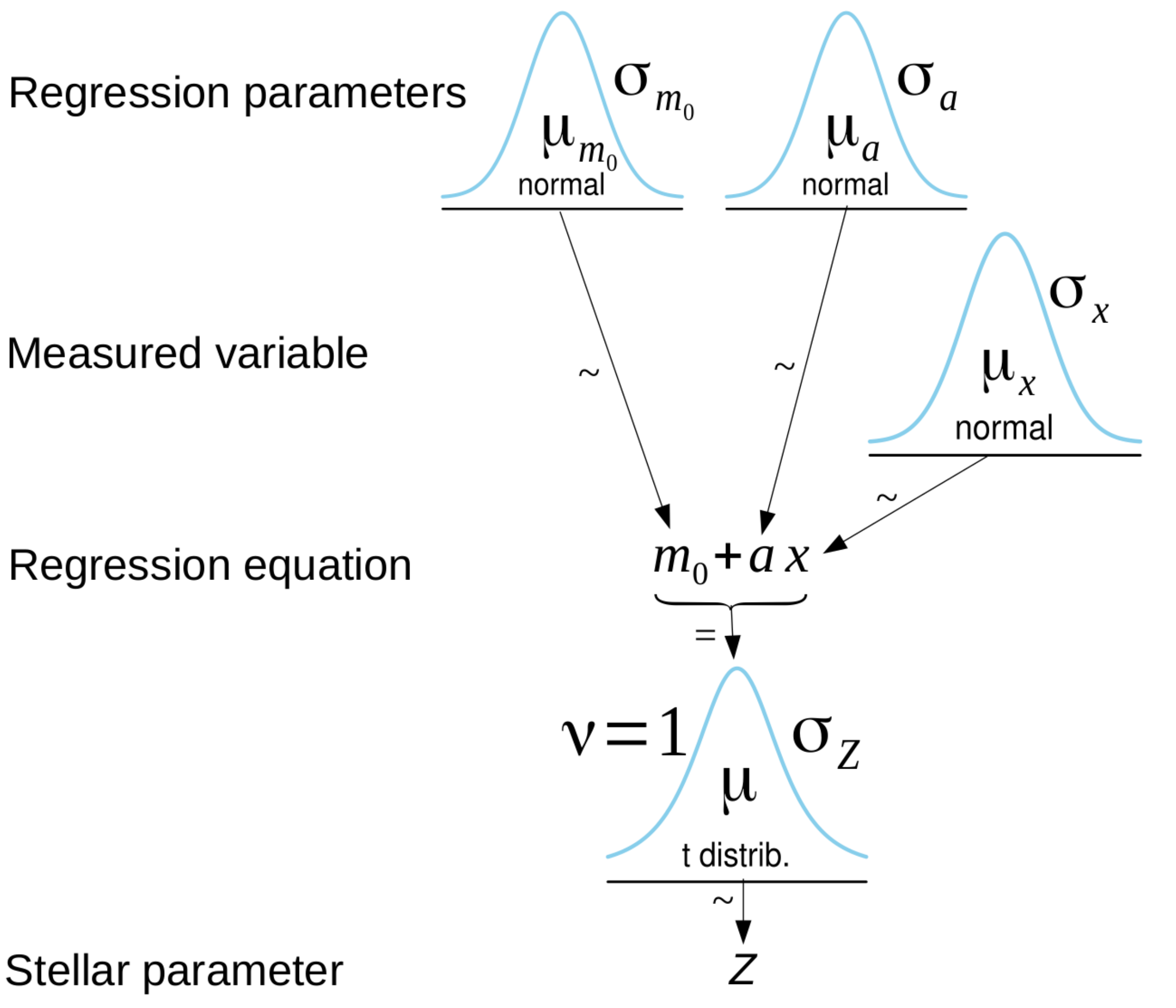}
	\caption{Kruschke style diagram for the linear Bayesian model (equation \ref{equation:bayesian_linear_individualline}). The probabilistic distributions of model parameters are shown. Arrows with the sign $\sim$ denote stochastic dependencies, and arrows with the sign = represent deterministic dependencies. During the Bayesian model fit, each of the variables in the iteration will be sampled for this distribution. For quadratic and cubic Bayesian models, the model framework is the same, but the regression formula has to be updated accordingly.}
	\label{Fig:krusche_style_diagrams_LinearModel}
\end{figure}

To establish empirical relations, we carried out Bayesian analysis between stellar parameters and EWs of spectral features. The Bayesian analysis is based on Bayes' theorem, which comes straight from the conditional probability. The theorem can be represented as 
\[~~~~~~~~~~~~~~~~~~~~~\ P(\theta \mid D) = \frac{P(D \mid \theta) P(\theta)}{P(D)}, \]
where $D$ is the measured EWs, and $\theta$ is the model parameters, such as $T_{eff}$ and $[Fe/H]$. $P(\theta)$ represents the priors on the model parameters. $P(D)$ is the probability of the observed data for normalisation. To define a descriptive model for the Bayesian analysis, we considered regression equations (linear or quadratic or cubic), which can be denoted mathematically as follows.
\begin{equation} \label{equation:linear_individualline}
\hat{z} = m0 + a\times x
\end{equation}
for the linear regression of a individual spectral feature,
\begin{equation} \label{equation:linear_combinedlines}
\hat{z} = m0 + a\times x + b\times y
\end{equation}
for the linear regression of a combination of two spectral features,
\begin{equation} \label{equation:quadratic_individualline}
\hat{z} = m0 + a \times x +  c \times x^2 
\end{equation}
for the quadratic regression of a individual spectral feature,
\begin{equation} \label{equation:quadratic_combinedlines}
\hat{z} = m0 + a \times x +  b \times y + c \times x^2 + d \times y^2 
\end{equation}
for the quadratic regression of two spectral features, and
\begin{equation} \label{equation:cubic_individualline}
\hat{z} = m0 + a \times x +  c \times x^2 + e \times x^3 
\end{equation}
for the cubic regression of an individual spectral feature, where $\hat{z}$ represents the predicted $T_{eff}$ or $[Fe/H]$, $x$ and $y$ are EWs of spectral features, and m0, a, b, c, d, and e are the coefficients of the regression equation. To consider the random variation of actual $z$ around the predicted $\hat{z}$ and to make the inference robust against outlier due to stellar contamination, we assumed that actual $z$ are distributed randomly according to a Student's t-distribution around the predicted $\hat{z}$ and with standard deviation $\sigma_z$ as
\[~~~~~~~~~~~~~~~~~~~~~~~~~~~~~~~~ z \sim \mathcal{T}(\hat{z}, \sigma^2_z). \]
The normality parameter, $\nu$ is 1. Thus, the probabilistic reformulation of all regression equations to model the data are
\begin{equation} \label{equation:bayesian_linear_individualline}
\begin{aligned}
z \sim {} & \mathcal{T}(m0 + a \times x,\,\sigma^2_z)\,,
\end{aligned}
\end{equation}
\begin{equation} \label{equation:bayesian_linear_combinedlines}
\begin{aligned}
z \sim {} & \mathcal{T}(m0 + a\times x + b\times y,\,\sigma^2_z)\,,
\end{aligned}
\end{equation}
\begin{equation} \label{equation:bayesian_quadratic_individualline}
\begin{aligned}
z \sim {} & \mathcal{T}(m0 +  a \times x +  c \times x^2,\,\sigma^2_z)\,,
\end{aligned}
\end{equation}
\begin{equation} \label{equation:bayesian_quadratic_combinedlines}
\begin{aligned}
z \sim {} & \mathcal{T}(m0 + a \times x +  b \times y + c \times x^2 + d \times y^2,\,\sigma^2_z)\,,
\end{aligned}
\end{equation}
and
\begin{equation} \label{equation:bayesian_cubic_individualline}
\begin{aligned}
z \sim {} & \mathcal{T}(m0 + a \times x +  c \times x^2 + e \times x^3,\,\sigma^2_z)\,
\end{aligned}
\end{equation}
for equations \ref{equation:linear_individualline}, \ref{equation:linear_combinedlines}, \ref{equation:quadratic_individualline}, \ref{equation:quadratic_combinedlines}, and \ref{equation:cubic_individualline}, respectively, where $x_{i} \sim \mathcal{N}(EW_{x,i}, \sigma^2_{EW_{x,i}})$ and $y_{i} \sim \mathcal{N}(EW_{y,i}, \sigma^2_{EW_{y,i}})$. Hence, the measured EW quantities are also distributed according to the normal distribution defined by the measured mean and sigma. In this way, Bayesian analysis naturally enables to incorporate both the measurement errors in the EWs and stellar parameters in the regression problem and propagates all the uncertainties forward in a self-consistent manner to the final empirical relations. A Schematic view of this formalism is shown in Fig. \ref{Fig:krusche_style_diagrams_LinearModel}.

\subsubsection{Specifying prior distribution on parameters} \label{Section:Specifying prior distribution on parameters}
In Bayesian statistics, model parameters m0, a, b, c, d, and e should not be treated as fixed variables. In fact, they are also coming from distributions. However, prior distribution on the parameters should be specified in such a way that it has virtually no biasing effect on the resulting posterior distribution. We chose the non-informative (noncommittal and vague) priors that are distributed as normal distributions with the following mean ($\mu$) and standard deviation: m0 = 4800 K  $\pm$ 1500 K and a, b = $-$100 K $\pm$ 1500 K for $T_{eff}$ -- EWs relations, and m0, a, b, c, d, e = 1 dex $\pm$ 10 dex for [$Fe/H$] -- EWs relations. 

\begin{figure*}
	\centering
	\includegraphics[width=0.99\textwidth,angle=0]{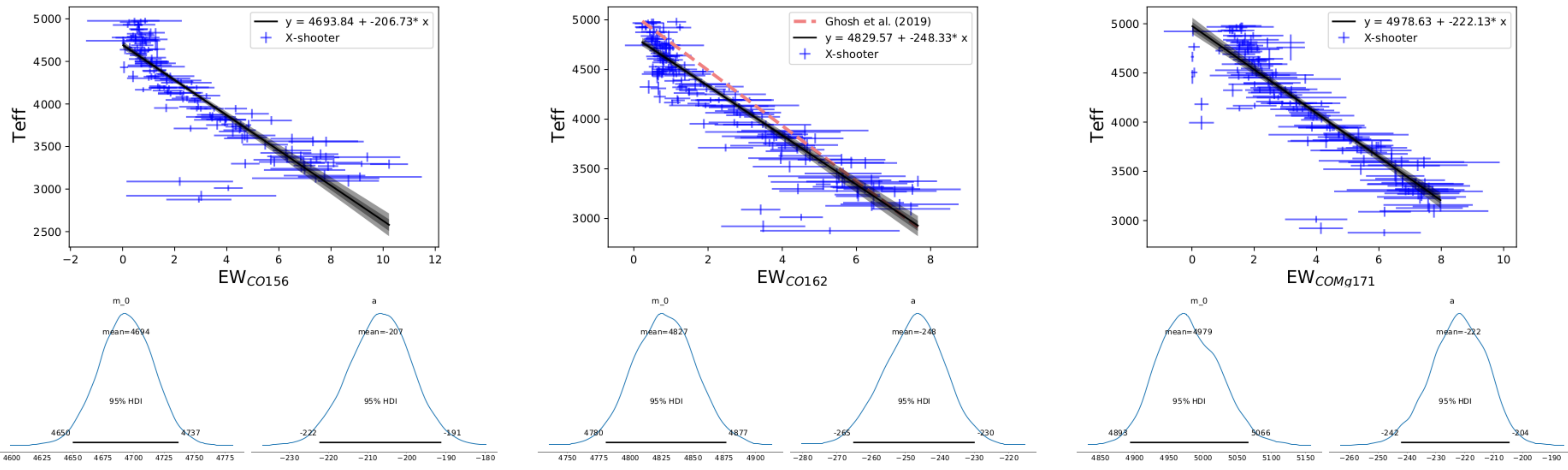} 
	\caption{Figure top panel displays the Bayesian fit of the linear model in $T_{eff}$ versus EWs (left: CO156, middle: CO162, right: COMg171). The black line shows the best fit, and the grey shaded regions represent 1 and 2 sigma intervals of the fitted model. Posterior distributions of the coefficients in the model are shown at the bottom panel. The 95\% Highest Density Interval (HDI) is also marked inside the posterior distributions. The summary of the posterior distributions and BIC are listed in Table \ref{tab:BayesianModelForTeff}. The red dashed line denotes established $T_{eff}$ versus EW$_{CO162}$ relationship of \citep{2019MNRAS.484.4619G} using 107 solar-neighbourhood stars.}
	\label{Fig:TeffEWCO162_fits}
\end{figure*}

\begin{table*}
	\centering
	\caption{Posteriors distribution of Bayesian models for $T_{eff}$ versus EWs of spectral features relationship and  comparison between various $T_{eff}$ correlations.}
	\label{tab:BayesianModelForTeff}
	\resizebox{0.99\textwidth}{!}{
		\begin{tabular}{ccrcccrrccc} 
			\hline
			\hline
			Index & T & N & m0$^a$ & a$^a$ & BIC & SEE & N$_{2\sigma clip}$ & SEE$_{2\sigma clip}$ & Relation$^a$ & Range$_{2\sigma clip}$$^b$ \\	
			\hline
			
			Mg150 & 177 & 148  & 5513 $\pm$ 121  & $-$648 $\pm$ 51  & 2289 & 432 & 134 & 332 & Equation (\ref{equation:bayesian_linear_individualline}) & [3089, 4978], [$-$1.70, 0.45] \\
			K152 & 177 & 138 & 4456 $\pm$ 47 & $-$955 $\pm$ 83 & 2117 & 366 & 128 & 314 & Equation (\ref{equation:bayesian_linear_individualline}) & [2921, 4933], [$-$1.70, 0.48] \\
			Fe153 & 177 & 141 & 4401 $\pm$ 41 & $-$699 $\pm$ 59  & 2163 & 340 & 122 & 249 & Equation (\ref{equation:bayesian_linear_individualline}) & [2876, 4871], [$-$2.35, 0.48] \\
			H155 & 177 & 154 & 4844 $\pm$ 47 & $-$491 $\pm$ 32  & 2294 & 334 & 139 & 254 &  Equation (\ref{equation:bayesian_linear_individualline}) & [3089, 4933], [$-$2.35, 0.45] \\
			CO156 & 177 & 152 & 4693 $\pm$ 22 & $-$207 $\pm$ 08  & 2137 &  272 & 132 & 153 & Equation (\ref{equation:bayesian_linear_individualline}) & [3096, 4962], [$-$2.35, 0.45] \\
			Mg157 & 177 & 173 &  4882 $\pm$ 46 & $-$203 $\pm$ 11  & 2590 & 368 & 153 & 260 & Equation (\ref{equation:bayesian_linear_individualline}) & [3143, 4980], [$-$2.35, 0.45] \\
			FeH158 & 177 & 176 & 4919 $\pm$ 34 & $-$489 $\pm$ 25 & 2600 & 345 & 154 & 220 & Equation (\ref{equation:bayesian_linear_individualline}) & [3127, 4977], [$-$2.35, 0.45] \\
			CO160 & 177 & 177 & 5052 $\pm$ 41 & $-$390 $\pm$ 23  & 2657 & 407 & 143 & 195 & Equation (\ref{equation:bayesian_linear_individualline}) & [3096, 4978], [$-$2.35, 0.45] \\
			Fe161 & 177 & 125 & 4444 $\pm$ 33 & $-$380 $\pm$ 24  & 1804 & 245 & 97 & 143 & Equation (\ref{equation:bayesian_linear_individualline}) & [2876, 4644], [$-$2.06, 0.45] \\
			CO162 & 177 & 175 & 4830 $\pm$ 24 & $-$248 $\pm$ 09  & 2436 & 209 & 154 & 123 & Equation (\ref{equation:bayesian_linear_individualline}) & [3096, 4968], [$-$2.35, 0.45] \\
			FeH162 & 177 & 93 & 4167 $\pm$ 34 & $-$283 $\pm$ 23  & 1377 & 328 & 71 & 131 & Equation (\ref{equation:bayesian_linear_individualline}) & [3096, 4431], [$-$1.78, 0.25] \\
			CO164 & 177 & 171 & 4814 $\pm$ 54 & $-$204 $\pm$ 13  & 2592 & 367 & 156 & 283 & Equation (\ref{equation:bayesian_linear_individualline}) & [3096, 4980], [$-$2.35, 0.45] \\
			CO166 & 177 & 172 & 4781 $\pm$ 30 & $-$441 $\pm$ 19 & 2488 & 293 & 152 & 168 & Equation (\ref{equation:bayesian_linear_individualline}) & [3096, 4978], [$-$2.34, 0.45] \\
			Al167 & 177 & 174 & 4785 $\pm$ 30 & $-$283 $\pm$ 12  & 2483 & 200 & 149 & 132 & Equation (\ref{equation:bayesian_linear_individualline}) & [3143, 4933], [$-$2.35 0.45] \\
			H168 & 177 & 148 & 3775 $\pm$ 43 & 909 $\pm$ 53  & 2171 & 333 & 116 & 141 & Equation (\ref{equation:bayesian_linear_individualline}) & [3558, 4978], [$-$1.70, 0.45] \\
			COMg171 & 177 & 156 & 4979 $\pm$ 45 & $-$222 $\pm$ 10  & 2208 & 239 & 128 & 107 & Equation (\ref{equation:bayesian_linear_individualline}) & [3096, 4933], [$-$1.70, 0.45] \\
			\hline
	\end{tabular}}
	T $-$ total nos. of data points; N $-$ no. of points having EWs > 0; N $_{2\sigma clip}$ $-$ no. of points after 2$\sigma$ clipping of N \\
	SEE $-$ standard error of estimate before 2$\sigma$ clipping; 
	SEE$_{2\sigma clip}$ $-$ standard error of estimate after 2$\sigma$ clipping.\\ 
	$^a$ Relation (equation) used to establish the correlation; m0 and a are coefficients of the equation.\\ $^b$ $T_{eff}$ and [$Fe/H$] range of the stars after 2$\sigma$ clipping.
	\\	
\end{table*}

\subsubsection{Model implementation}
We implemented our probabilistic models in PyMC3 for automatic Bayesian inference \citep{10.7717/peerj-cs.55}. PyMC3 uses a No-U-Turn Sampler (NUTS), a self-tuning variant of Hamiltonian Monte Carlo (HMC) to fit the model. Three independent sampling chains were run for 500 tunes and 2000 draw iterations. The posterior distributions of model parameters were then investigated to study the credibility across parameter values. 

\subsubsection{Posterior distribution and temperature indicators} \label{Section:Teff-EWs_EmpiricalRelations}
For new empirical relations, Mg150, K152, Fe153, H155, CO156, Mg157, FeH158, CO160, Fe161, CO162, FeH162, CO164, CO166, Al167, H168, and COMg171 lines were undertaken. We first adopted the linear model of equation \ref{equation:bayesian_linear_individualline} to fit the linear relationship between $T_{eff}$ and EWs of spectral features. Using the values of priors of model parameters as previously mentioned, the data were fitted in the Bayesian framework. Three independent sampling chains for the individual parameters converged well to the same posterior distribution for all spectral features. Fig. \ref{Fig:TeffEWCO162_fits} shows the linear Bayesian fit of the $T_{eff}$ versus EWs relation for CO162, Al167, and COMg171. The blue symbols refer to the data used for the Bayesian fit. Blackline is the best fit, and the grey shaded regions are the 1 and 2 sigma intervals of the fitted model. The bottom panel of Fig. \ref{Fig:TeffEWCO162_fits} displays the posterior distribution of model parameters. The 95\% Highest Density Interval (HDI) is also marked by the black bar on the floor of the posterior distributions in Fig. \ref{Fig:TeffEWCO162_fits} which yields parameter values of a total probability of 95\%. Fig. \ref{Fig:TeffEWCO162_fits} also shows that the most credible parameters mimic the data very well. It is to note that we provide figures for linear fit and posterior distribution for other features as supplementary material for the readers' convenience. The summary of the posterior distributions and Bayesian information criteria (BIC) are listed in Table \ref{tab:BayesianModelForTeff} for all spectral features. In addition, Table \ref{tab:BayesianModelForTeff} gives the standard error of estimate before and after the 2$\sigma$ clipping (SEE and SEE$_{2\sigma clip}$, respectively), the no. of points remaining after 2$\sigma$ clipping (N$_{2\sigma clip}$) and the range of $T_{eff}$ and [$Fe/H$] after the 2$\sigma$ clipping (Range$_{2\sigma clip}$) for various relations. It should be noted that 2$\sigma$ clipped data were only used for the estimation of SEE.

\begin{figure}
	\centering
	\includegraphics[scale=0.44]{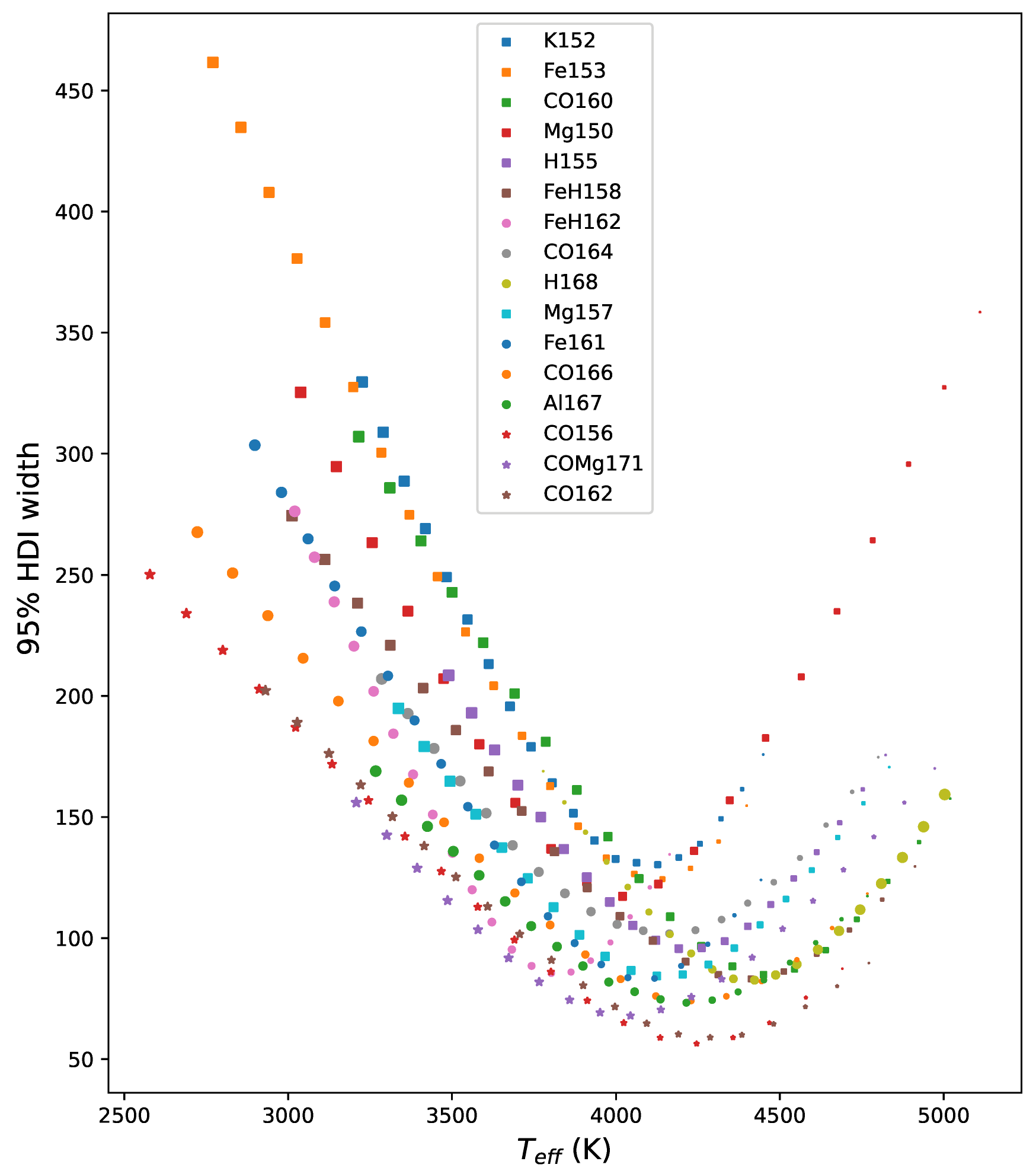}
	\caption{Comparison between 95\% HDI widths of the model predicted $T_{eff}$ versus $T_{eff}$s for various spectral feature. They are derived using posterir distribution of the fitted m0 and a parameters, and  evaluvated at an uniform grid of EWs for each model. Smaller the HDI width, higher the precision of the model's prediciton. The legends are shown in the increasing order of the precision of model's prediction. The `*' symbols represents the best three models. The size of the symbols corresponds to the value of EWs in each case. Uncertanity from the EW measurment is not considered here.}
	\label{Fig:best_linearmodel_for_Teff}
\end{figure}

As can be seen from Table \ref{tab:BayesianModelForTeff}, several spectral features such as CO156, CO160, Fe161, CO162, FeH162, CO166, Al167, H168, and COMg171 are good indicators of $T_{eff}$ with a SEE$_{2\sigma clip}$ less than 200 K. 
 The posterior distribution of the fitted model parameters in the bayesian models also provide us an estimate of the precision by which each model predicts the $T_{eff}$ given a precise EW measurement. We calculated the 95\% HDI width of the $T_{eff}$ predicted by each model for a uniform grid of EW values. A higher precision model will have smaller HDI for the $T_{eff}$ prediction. Fig. ~\ref{Fig:best_linearmodel_for_Teff} shows these HDI estimates for differnt models as a function of the stellar temperature. The legends in the plot are ranked in the increasing order of the model's $T_{eff}$ precision. Uncertainty from the EW measurement is not considered in this comparison. 
It is evident from Fig.~\ref{Fig:best_linearmodel_for_Teff} that both CO156 and CO162 are the best $T_{eff}$ predictors at $T_{eff}$s $>$ 4000 K, however, at $T_{eff}$s $<$ 4000 K, CO156, CO162 and COMg171 provide the best prediction. Thus, we validated CO156 and CO162 above 4000 K (to 5000 K), and CO156, CO162, and COMg171 below 4000 K for precise estimation of $T_{eff}$ for cool giants. The typical accuracies of these three models are 153 K, 123 K and 107 K, respectively. Depending on the EW measurement precision and stellar temperature, this figure can be used to select a set of most useful line models for $T_{eff}$ calculation.

Furthermore, we compared our $T_{eff}$--EW$_{CO162}$ relationship with that of \citet{2019MNRAS.484.4619G} (red dashed line in Fig. \ref{Fig:TeffEWCO162_fits}), which was established using 107  solar-neighbourhood giants. A significant difference between two relationship is apparent at $T_{eff}$ $>$ 3500 K. The $T_{eff}$ tends to be overestimated by up to $\sim$ 210 K if empirical relation of \citet{2019MNRAS.484.4619G} is adopted for $T_{eff}$ estimation. The difference could be because of the systematics between the two methods used to measure atmospheric parameters of the sample giants or the large metallicity range of our sample. 

\begin{figure*}
	\centering
	\includegraphics[scale = 0.3]{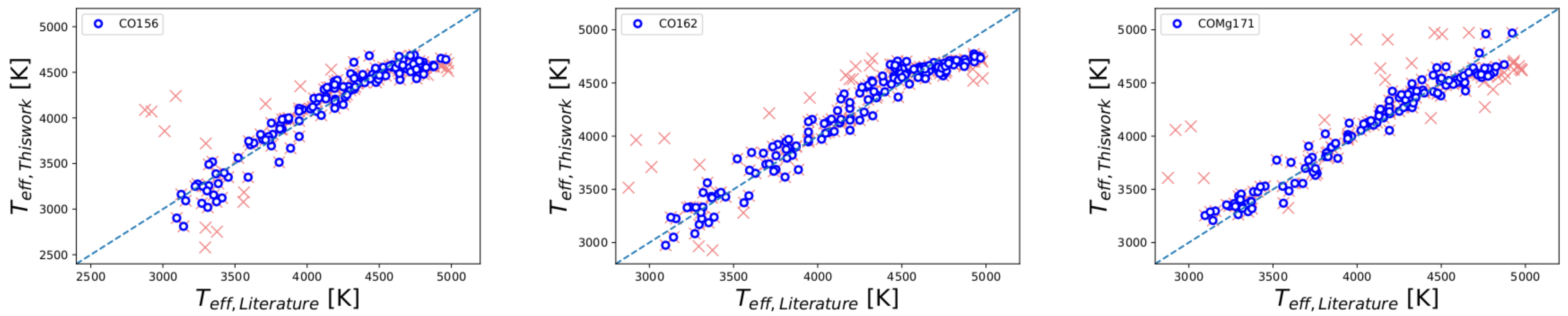} 
	\caption{$T_{eff}$ comparison between derived values from the empirical relations and literature values (left: CO156, middle: CO162, right: COMg171). The coefficients of empirical relations are listed in Table \ref{tab:BayesianModelForTeff}. The dashed line displays the one-to-one correspondence of $T_{eff}$. The red cross symbols represent all of our samples and the blue circles show stars after 2$\sigma$ clipping. It is important to note that we removed 2$\sigma$ outliers for the estimation of SEE only.}		
	\label{Fig:Teff_fit_residual}
\end{figure*}

The linear empirical solutions between $T_{eff}$ and CO156, CO162, and COMg171 were then applied to each star (excluding outliers) in the sample, and the resulting $T_{eff}$ values were compared to the literature value as displayed in Fig. \ref{Fig:Teff_fit_residual}. The mean and standard deviation of the fit residuals are $\Delta T_{eff, Avg}$ = 24 K and $\sigma_{T_{eff}}$ = 151 K for CO156, $\Delta T_{eff, Avg}$ = $-$9 K and $\sigma_{T_{eff}}$ = 123 K for CO162, and $\Delta T_{eff, Avg}$ = $-$11 K and $\sigma_{T_{eff}}$ = 106 K for COMg171, respectively. 

 \begin{figure}
	\includegraphics[width=0.47\textwidth,angle=0]{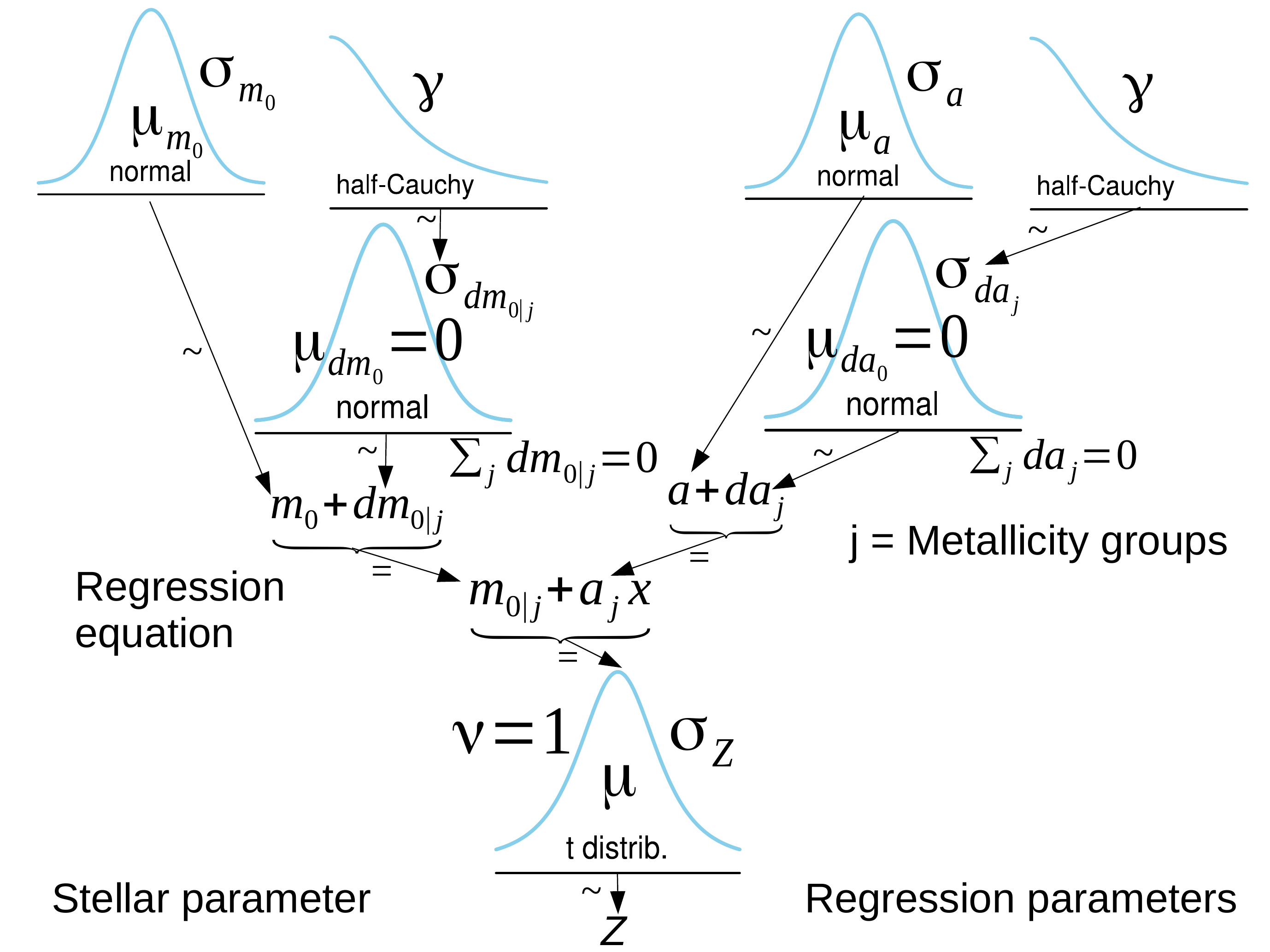}
	\caption{Kruschke style diagram of the Hierarchical Bayesian model to study possible metallicity dependence on $T_{eff}$ versus EWs relationship. The coefficients of the linear relationship are hierarchically split into a group average plus delta differences for each metallicity group. The hyper parameters determine the scatter in these coefficients across the metallicity groups. See the text for details.}
	\label{Fig:KruschkeDiagramHBM}
\end{figure}

 \begin{figure*}
	\centering
	\includegraphics[width=1.0\textwidth,angle=0]{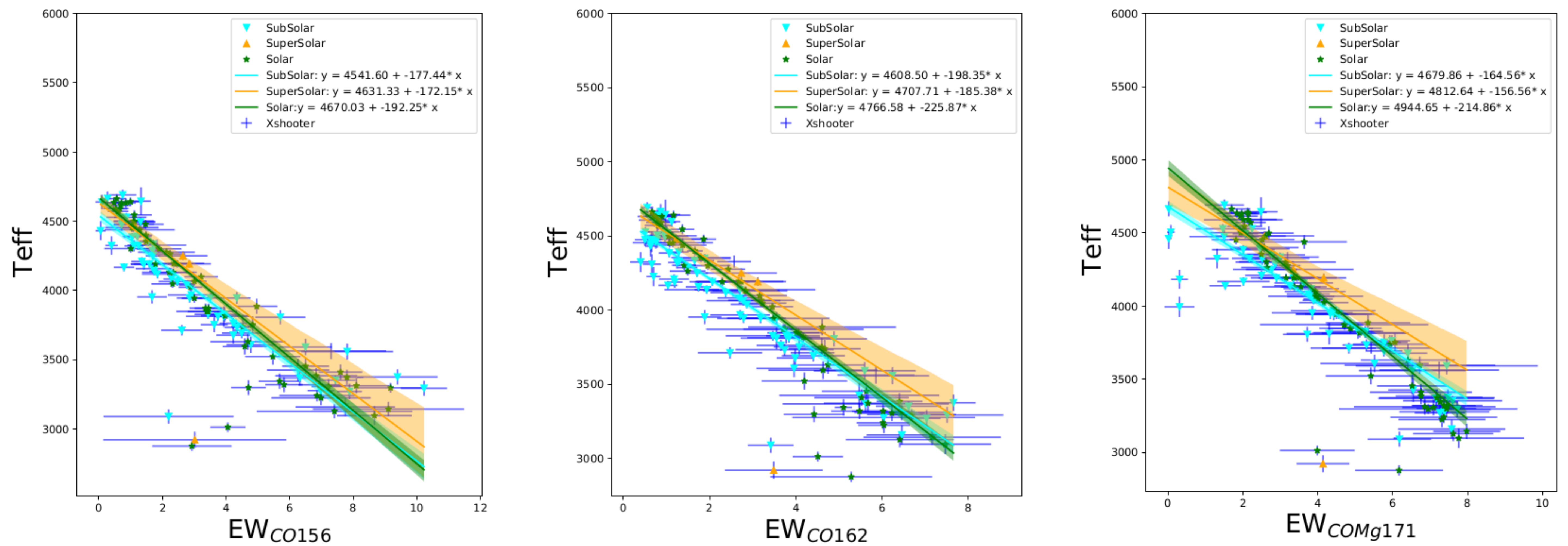}
	\caption{Hierarchical Bayesian model for investigating the metallicity effect in $T_{eff}$ versus EWs relationship by grouping data into three separate metallicity groups. It is to note that we adopted here stars of $T_{eff}$ $\leq$ 4700 K. The cyan color corresponds to subsolar ([$Fe/H$] $<$ $-$0.3 dex) group, the orange color represents the supersolar ([$Fe/H$] $>$ 0.3 dex) group, and the green color is for solar metallicity ($-$ 0.3 dex $<$ [$Fe/H$] $<$ + 0.3 dex) group. All the solid lines are the best fit and the color shaded regions are the 1$\sigma$ interval of the fitted models for three different metallicity groups.}
	\label{Figure:Hierarchical_Bayesian_Linear_model_fit}
\end{figure*}

\begin{figure*}
	\centering
	\includegraphics[width=1.0\textwidth,angle=0]{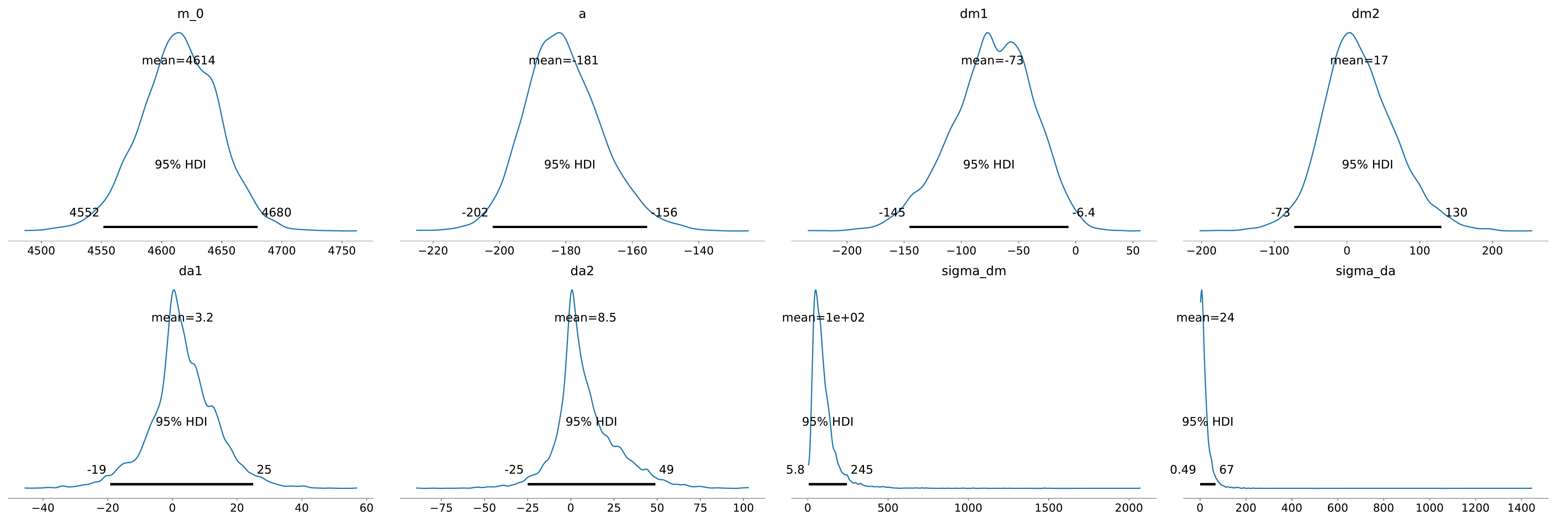} \\
	\includegraphics[width=1.0\textwidth,angle=0]{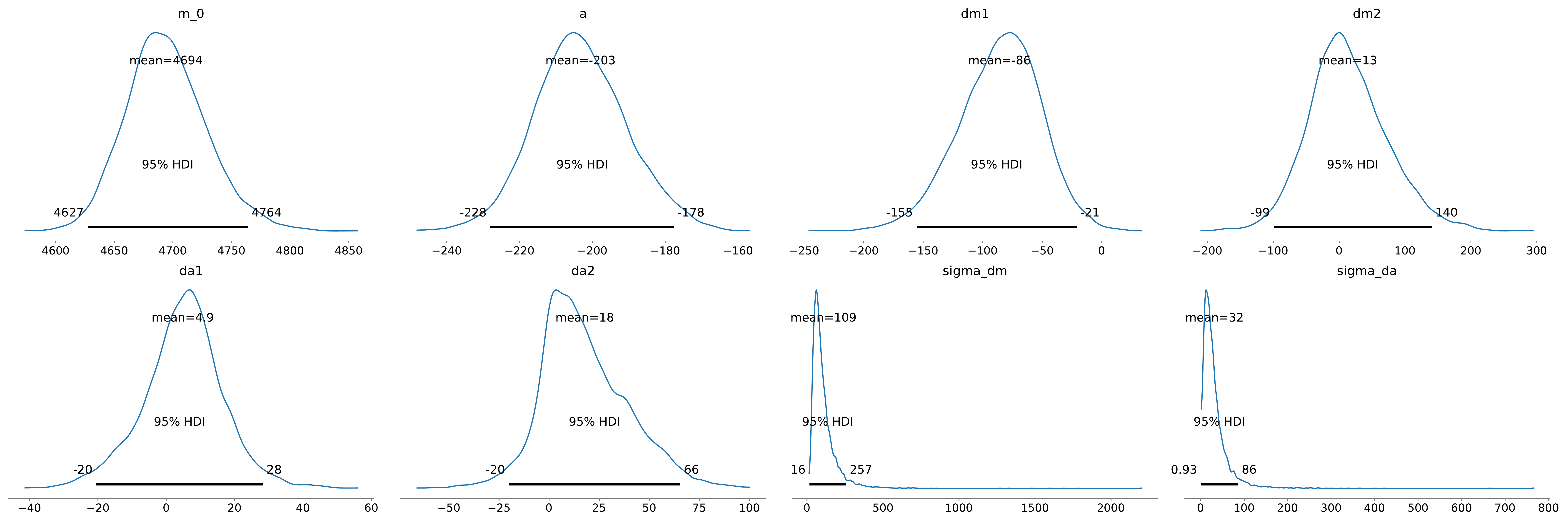} \\
	\includegraphics[width=1.0\textwidth,angle=0]{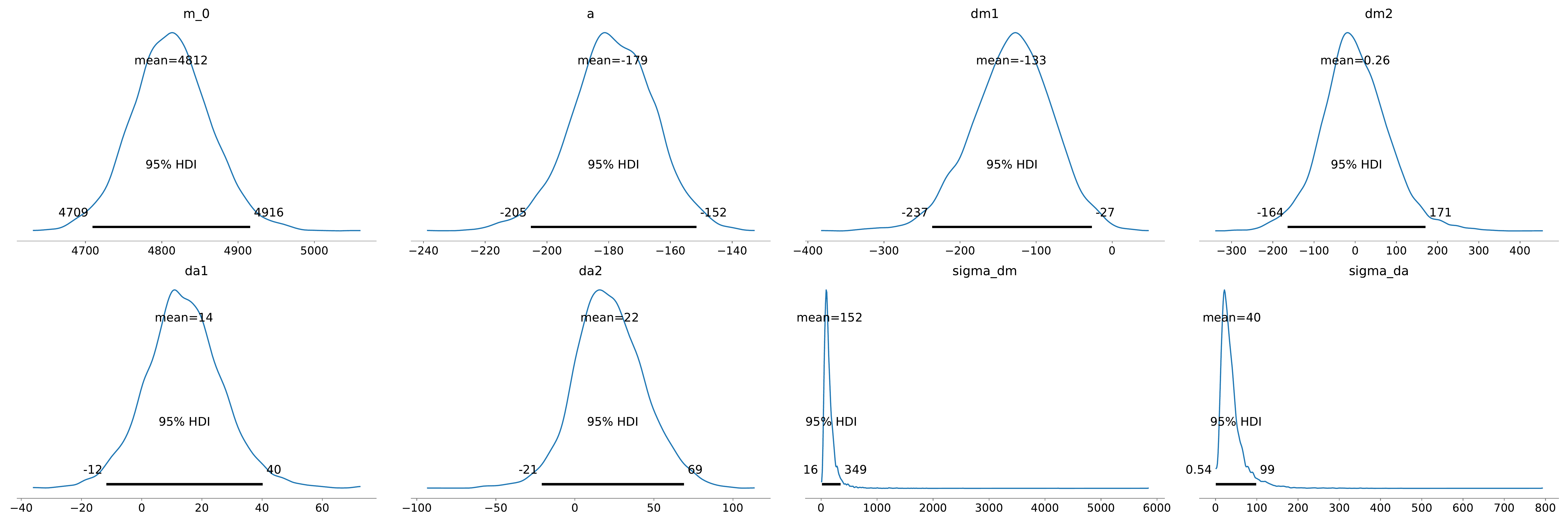}
	\caption{The top two panels represent the posterior distribution of the group average coefficients for the linear model and the group-specific difference to the coefficients of the first two metallicity groups for CO156. The third metallicity group's value is given by the constrain that the sum of the group differences should be equal to zero. Panels 3 and 4 show the same posterior distribution as in the case of CO156, but for CO162. The bottom two panels are for COMg171.}
	\label{Figure:Hierarchical_Bayesian_Linear_model_posterior}
\end{figure*}

From Fig.~\ref{Fig:TeffEWCO162_fits}, it is evident that the EWs and $T_{eff}$ relationships do not follow a similar linear trend at $T_{eff}$ > 4700 K. Also, our estimated $T_{eff}$s show a deviation from the literature values above the same temperature (see Fig.~\ref{Fig:Teff_fit_residual}). It is expected as spectral features become very weak at $T_{eff}$ > 4700 K (see, Fig~\ref{Fig:index-Teff_behaviour}), and we have adopted a linear model for the convergence of the posterior for all indices, avoiding significant degeneracy and the simpler interpretation. Thus, for precise estimation in temperature, the established linear relationships are valid at $T_{eff}$ $\leq$ 4700 K.

Furthermore, four outliers stars (SHV 0549503-704331, HV 12149, HV 2446 and OGLEII DIA BUL-SC30 0707) can be seen clearly in Fig.~\ref{Fig:TeffEWCO162_fits}, and these stars could be variable stars. As discussed earlier, we have removed the known Mira and OH/IR variables from our sample using SIMBAD, however, it is possible that some variable stars can be contaminating our sample according to the other catalogues. We did not play cat and mouse with our data and instead tried to make our analysis robust to the outliers. Even though we found that the first three outlier stars according to \citet{2018A&A...616A...1G} and the last one based on \citet{2005A&A...443..143G} are Mira variables.
	
\subsubsection{Metallicity effect on $T_{eff}$--EWs relations} \label{Section:MetaEffectOnTeff-EWs}
In section \ref{Section:Teff-EWs_EmpiricalRelations}, we established empirical relations assuming that the EWs depend only on the $T_{eff}$. However, the line strength of a spectral feature is not completely insensitive to [$Fe/H$]. Moreover, some sort of metallicity dependence on EWs estimation can be visualized from Fig. \ref{Fig:index-meta_behaviour} especially for CO156, CO162, and COMg171, the most favourable $T_{eff}$ predictors. Hence, the influence of metallicity on the derived empirical $T_{eff}$--EWs relations are needed to be investigated. As discussed in the previous section, CO156, CO162, and COMg171 behave linearly with $T_{eff}$ at $T_{eff}$ $<=$ 4700 K, and thus, we considered a linear model and data points at $T_{eff}$ $<=$ 4700 K for this study.

The multiple models for each metallicity group can be developed for the analysis. However, modelling a heterogeneous set of disjoint metallicity groups individually might not produce reliable results because of a few giants identified in super-solar and sub-solar groups. Thus, for this analysis, we constructed a Hierarchical modelling or multilevel modelling, which allows us to simultaneously model different metallicity groups. In Hierarchical modelling, as the parameters are nested within one another at multiple levels, the model pools information across the groups while fitting. 

In performing the Bayesian analysis, we modelled the coefficients of the linear regression, equation \ref{equation:bayesian_linear_individualline}, as a sum of a group average plus a delta specific to each metallicity group. The delta correction term to each group is also distributed as normal distributions with the mean zero and finite sigma. This sigma which represents the scatter in the metallicity group differences is hierarchically sampled from a half-Cauchy distribution with hyper-parameters. Thus, the reformulation of the regression equation \ref{equation:bayesian_linear_individualline} is
\begin{equation} \label{equation:Hierarchical_bayesian_linear_individualline}
\begin{aligned}
z \sim {} & \mathcal{T}((m0 + dm0_j) + (a + da_j) \times x,\,\sigma^2_z)\,.
\end{aligned}
\end{equation}

 The Kruschke style diagram of our Hierarchical Bayesian model is shown in Fig \ref{Fig:KruschkeDiagramHBM}. The Bayesian fits and three different relations along with their confidence for three different metallicity groups for CO156, CO162, and COMg171 are shown in Fig. \ref{Figure:Hierarchical_Bayesian_Linear_model_fit}, and the posterior distribution of model parameters is in Fig. \ref{Figure:Hierarchical_Bayesian_Linear_model_posterior}. Comparing the posterior distributions of the group differences (dm and da term) and the confidence interval for three metallicity groups (as shown in Fig. \ref{Figure:Hierarchical_Bayesian_Linear_model_fit}) together we found that the solar metallicity group's $T_{eff}$--CO156 relation is significantly different from the subsolar metallicity group at $T_{eff}$ $\geq$ 3500 K. The effective temperature tends to be overestimated by up to $\sim$ 130 K if we estimate $T_{eff}$ for the subsolar metallicity group's stars using the empirical relationship of solar metallicity group. For CO162, the significant difference is also evident between the solar and subsolar metallicity groups at $T_{eff}$ $\geq$ 3700 K, and the effective temperature tends to be underestimated by up to $\sim$ 150 K for the solar metallicity group in comparison to $T_{eff}$ estimated by empirical relations of subsolar metallicity group. For COMg171, the effective temperature tends to be overestimated by up to $\sim$ 260 K at $T_{eff}$ $\geq$ 3900 K, but rather underestimated by up to $\sim$ 120 K at $\leq$ 3500 K if we estimate $T_{eff}$ for the subsolar metallicity group’s stars using the empirical relationship of solar metallicity group. However, the difference between the solar and supersolar metallicity groups is not as statistically significant for all spectral features studied here. In addition, we estimated SEE$_{2\sigma clip}$ for three different metallicity groups as listed in Table~\ref{tab:HierarchicalBayesianmodelForSEE}.
 
\begin{center}
	\begin{table}
		\caption{Accuracy of Hierarchical Bayesian fit empirical relations for three different metallicity groups.}
		\label{tab:HierarchicalBayesianmodelForSEE}
		\resizebox{0.47\textwidth}{!}{
			\begin{tabular}{cccc} 
				\hline
				\hline
				Index & SEE$_{2\sigma clip}$ & SEE$_{2\sigma clip}$ & SEE$_{2\sigma clip}$ \\
				& (Solar) & (Subsolar) & (Supersolar) \\
				\hline
				
				CO156 & 118 & 94& 56 \\
                CO162 & 56 & 117 & 46 \\
               COMg171 & 87 & 97 & 49  \\				
				\hline
		\end{tabular}}
	\end{table}
\end{center}

\begin{figure}
	\centering
	\includegraphics[width=0.45\textwidth,angle=0]{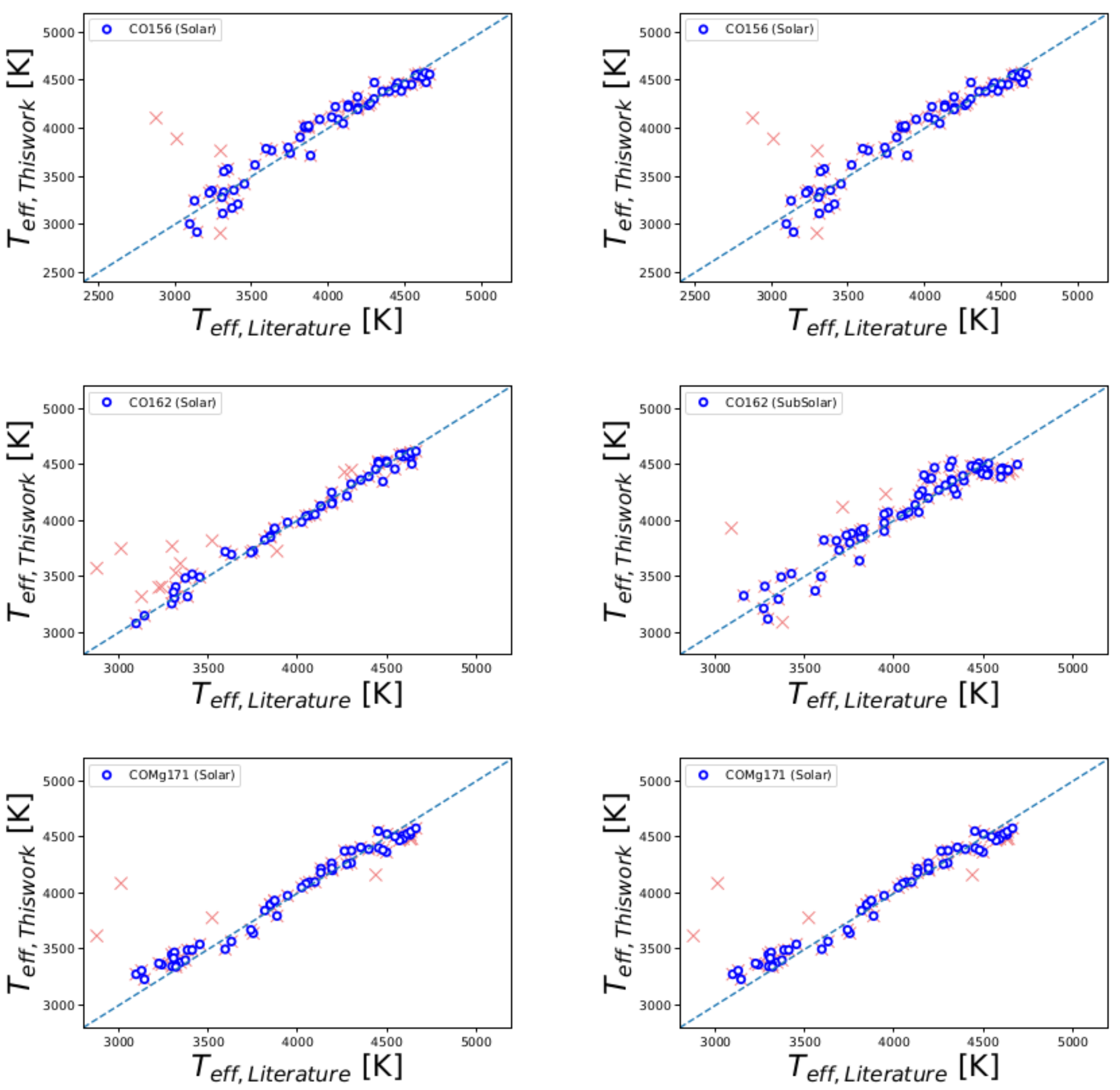}
	\caption{$T_{eff}$ comparison between derived values from the empirical relations (as given in the inset of Fig. \ref{Figure:Hierarchical_Bayesian_Linear_model_fit}) and literature values for solar and subsolar metallicity groups. The coloured `X' symbols display all stars in the respective metallicity group and blue circles represent stars after 2$\sigma$ clipping. The dashed line displays the one-to-one correspondence of $T_{eff}$. The supersolar group was not considered here as the difference in posterior distribution with the solar group is not as statistically significant. The top, middle, and bottom panels are for CO162, Al167, and COMg171, respectively. The left panel is for the solar metallicity group and the subsolar group is in the right panel.}
	\label{Figure:1to1_correspondence_for_Hierarchical Bayesian_model}
\end{figure}

\begin{figure} 
	\centering
	\includegraphics[scale = 0.45]{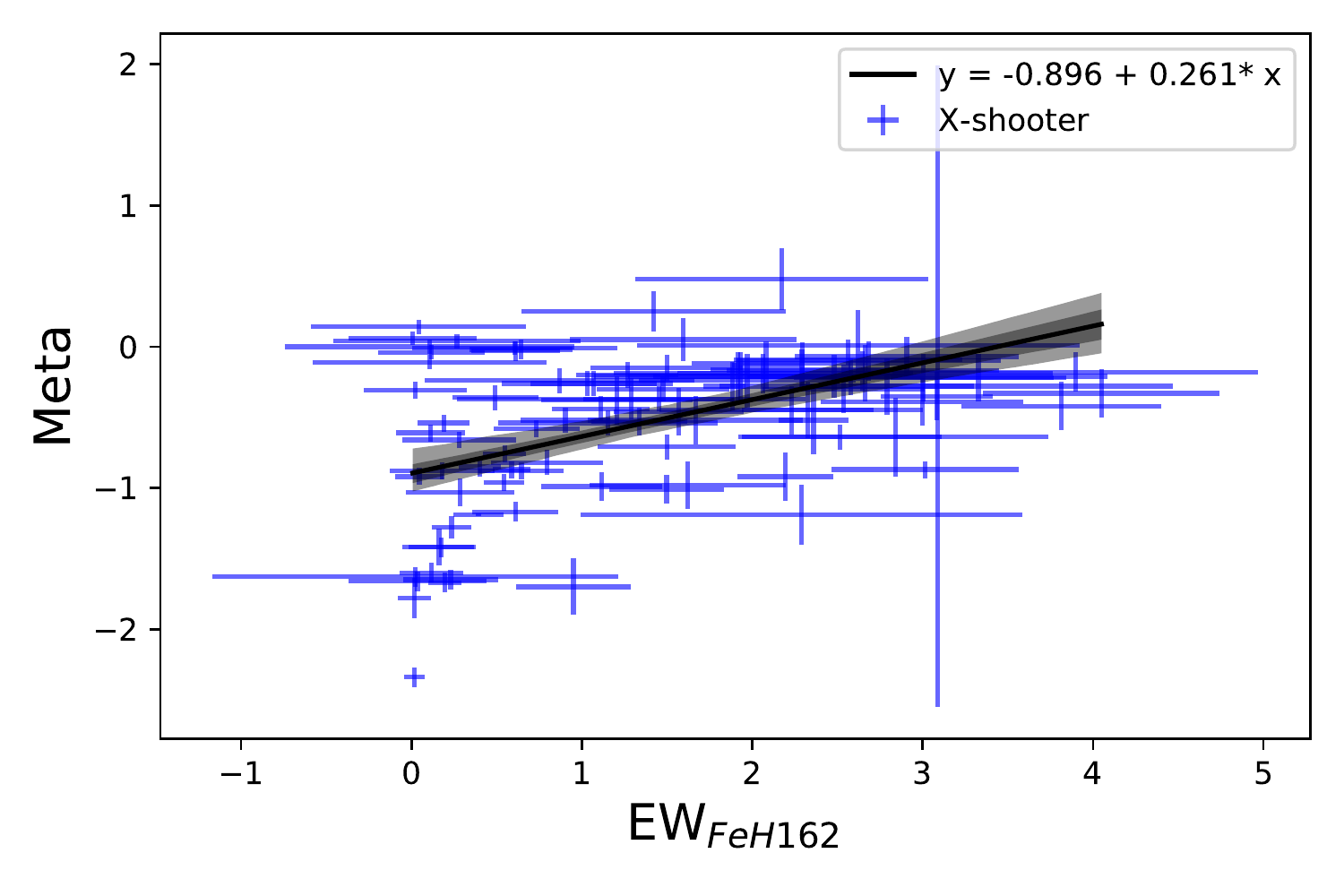} \\
	\includegraphics[scale = 0.16]{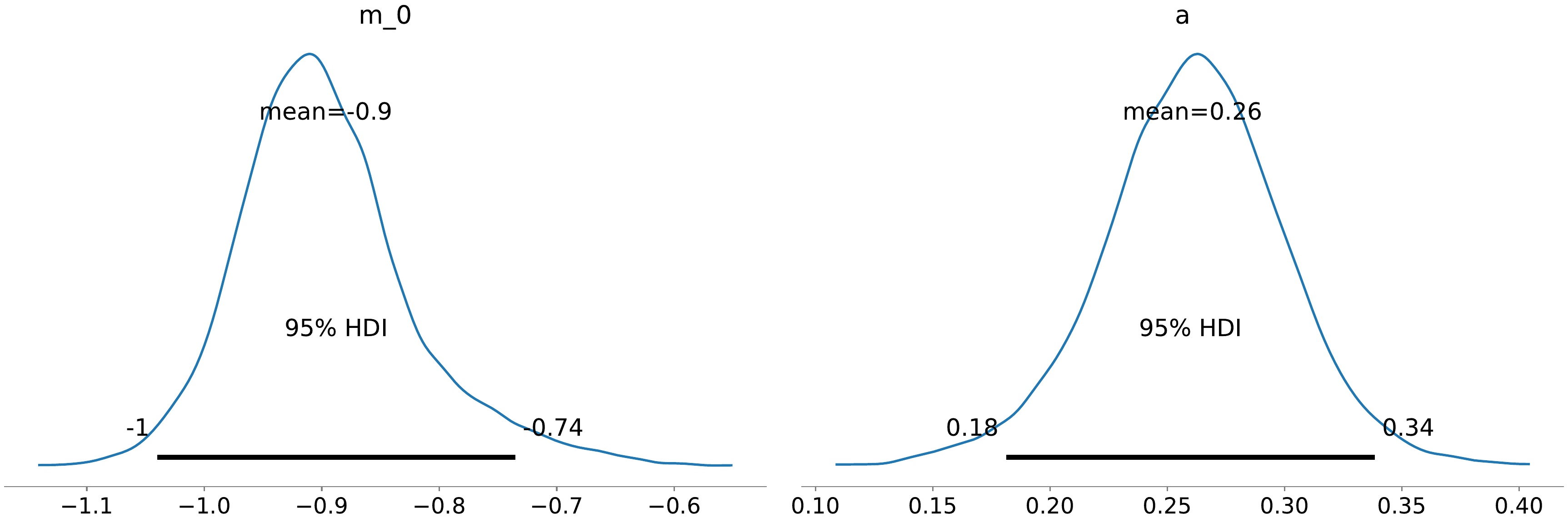} \\			
	\includegraphics[scale = 0.45]{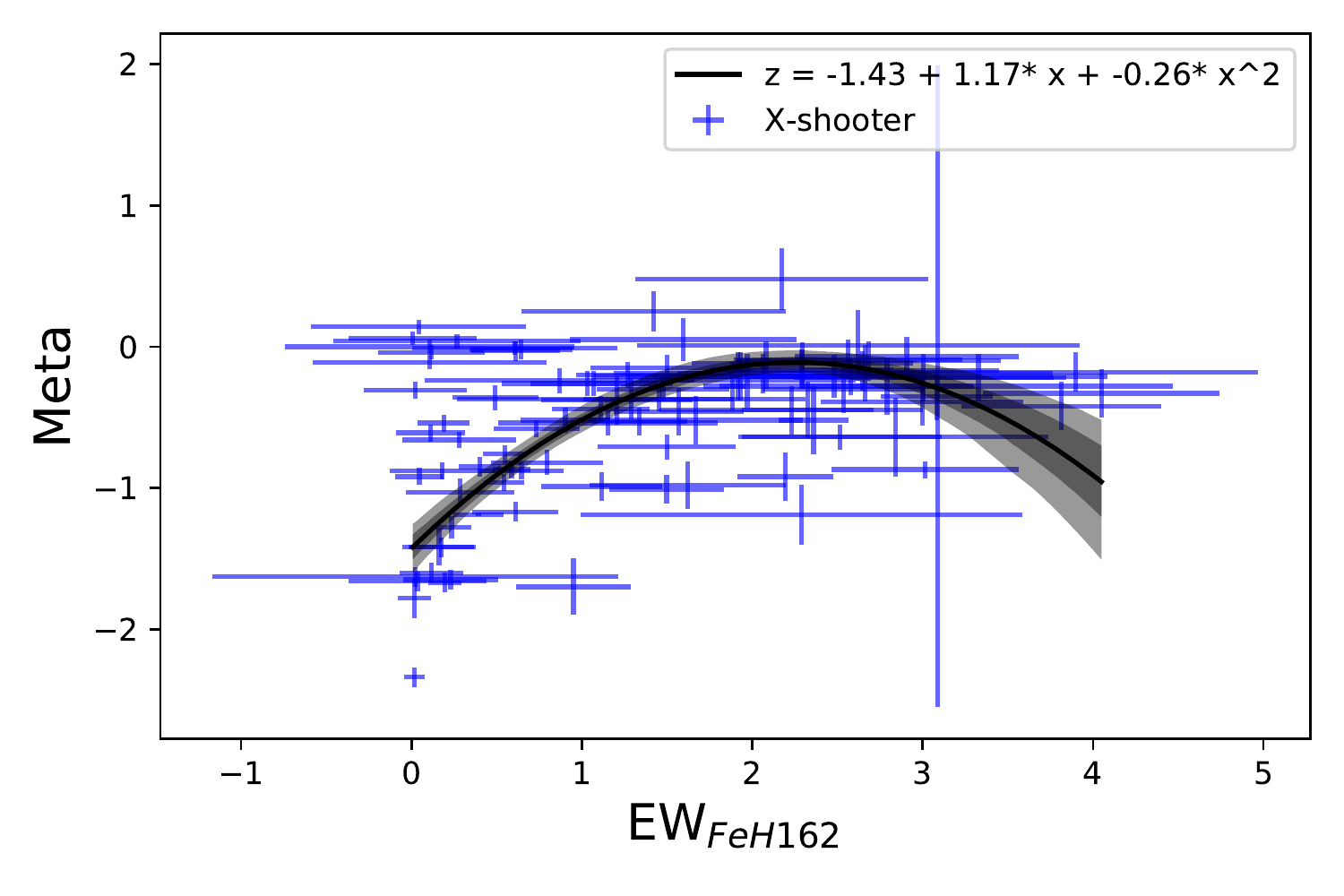} \\
	\includegraphics[scale = 0.15]{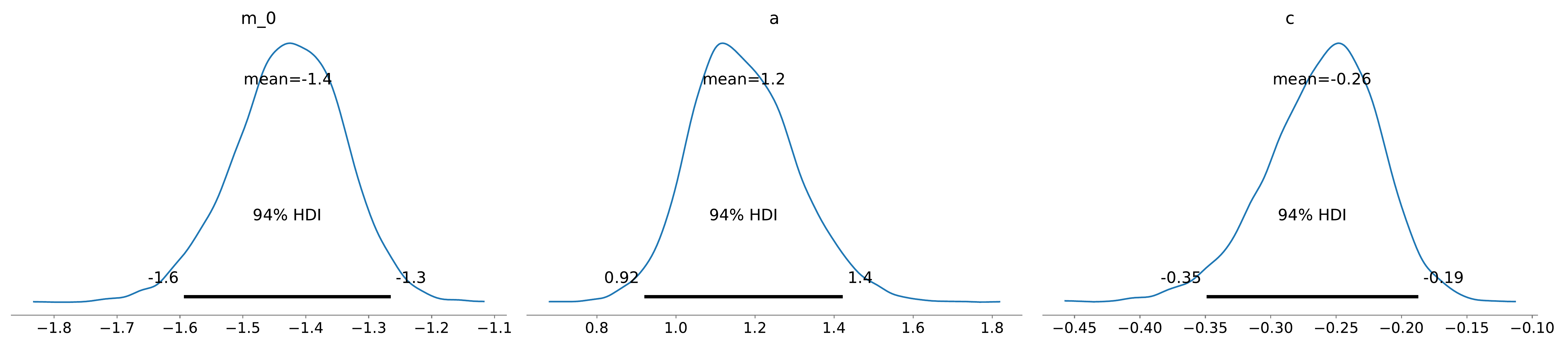} \\			 			
	\includegraphics[scale = 0.45]{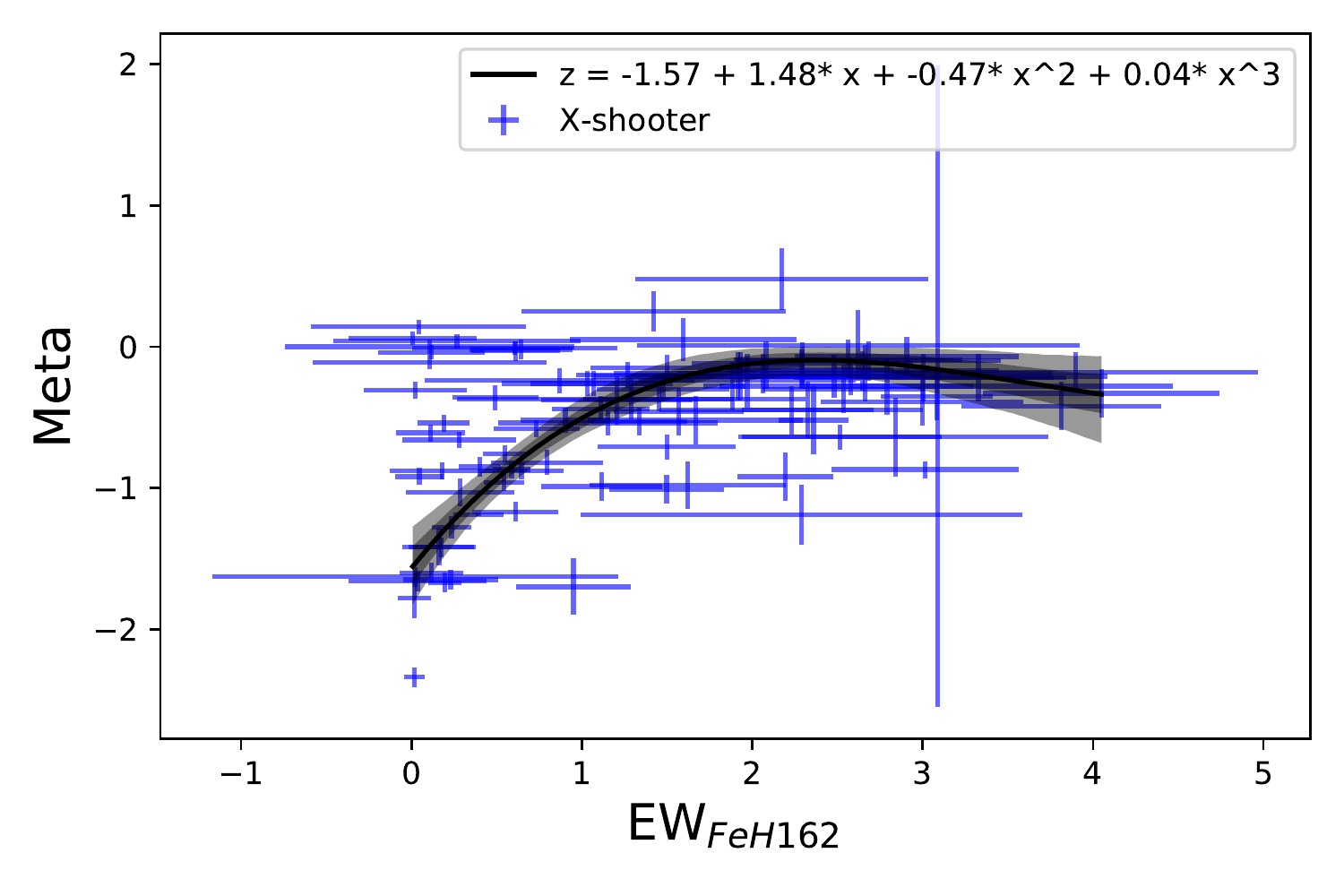} \\
	\includegraphics[scale = 0.12]{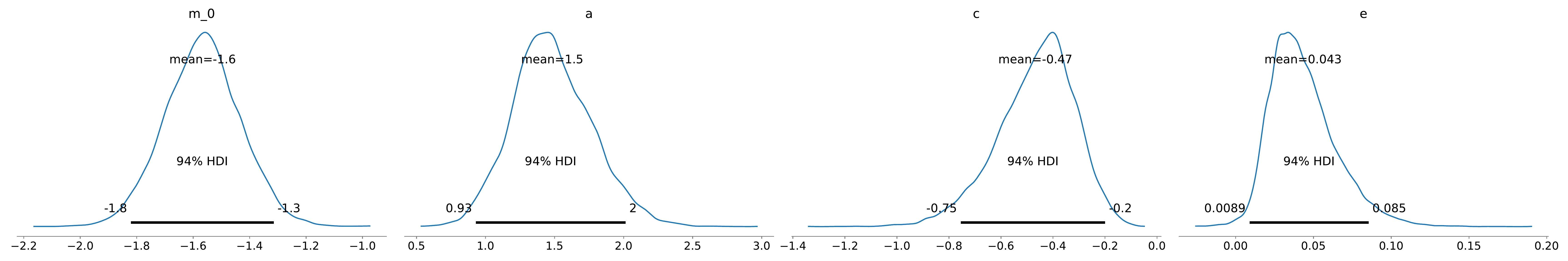} \\ 			
	\caption{Figure displays the Bayesian fit in [$Fe/H$] versus EW$_{FeH162}$. The best fit is shown by the black solid line, and the grey shaded regions represent 1 and 2 sigma intervals of the fitted model (from the top, panel 1: linear model,  panel 3: quadratic model, and panel 5: cubic model). Posterior distributions of the coefficients for various models are shown in the immediate bottom panel of the fitted model. The 95\% HDI is also marked inside the posterior distributions. The summary of the posterior distributions and BIC are tabulated in Table \ref{tab:BayesianModelForMetallicity}.} 
	\label{Figure:meta_FeH_BayesianFit} 
\end{figure}

\begin{figure} 
	\centering
	\includegraphics[scale = 0.45]{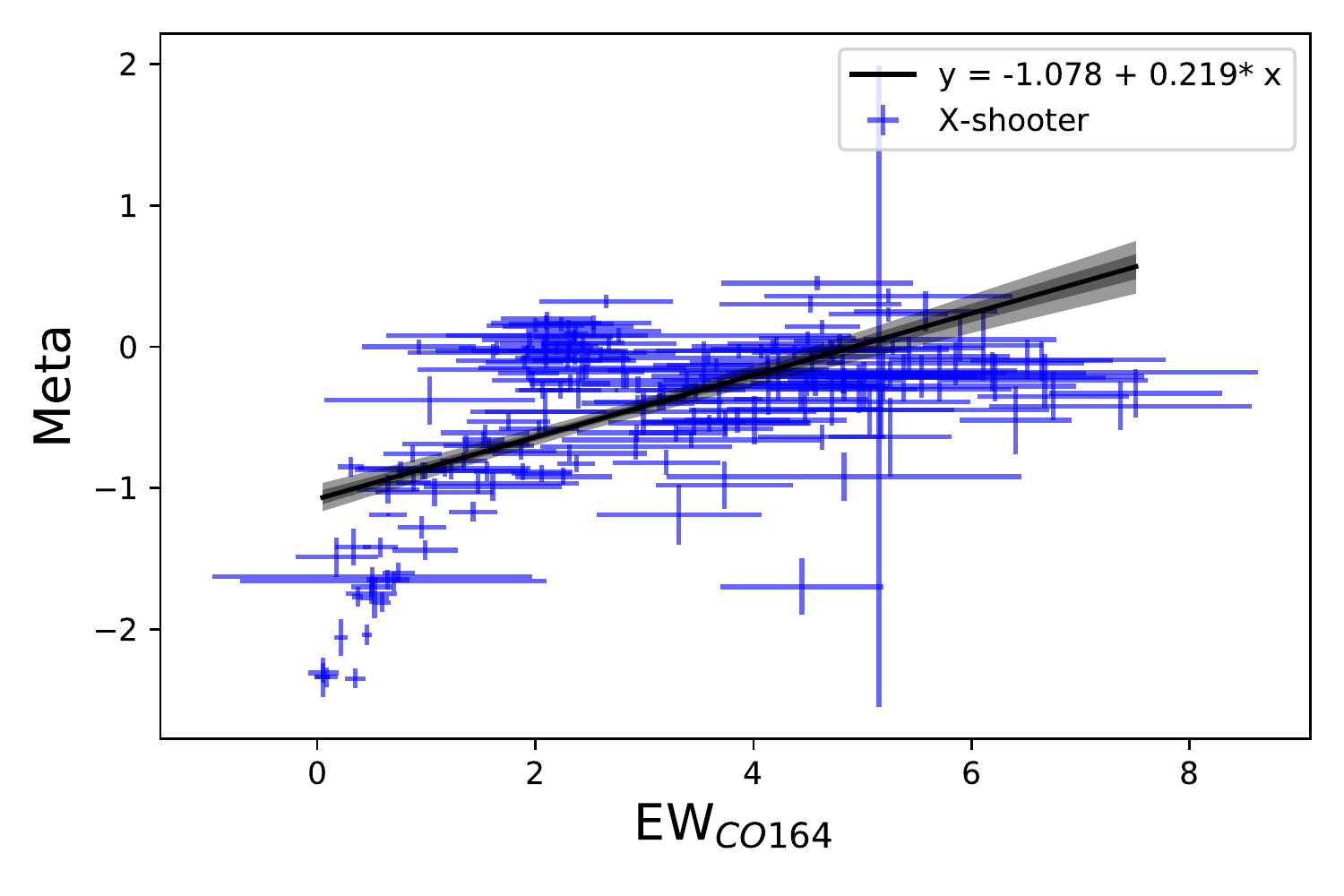} \\
	\includegraphics[scale = 0.16]{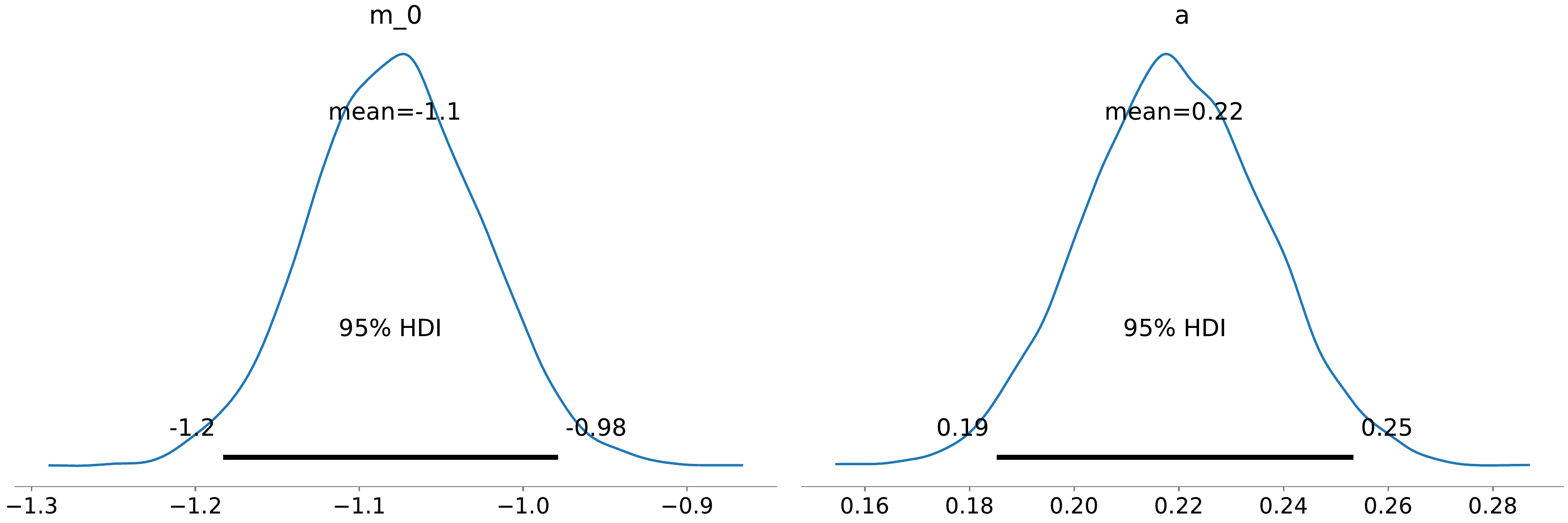} \\			
	\includegraphics[scale = 0.45]{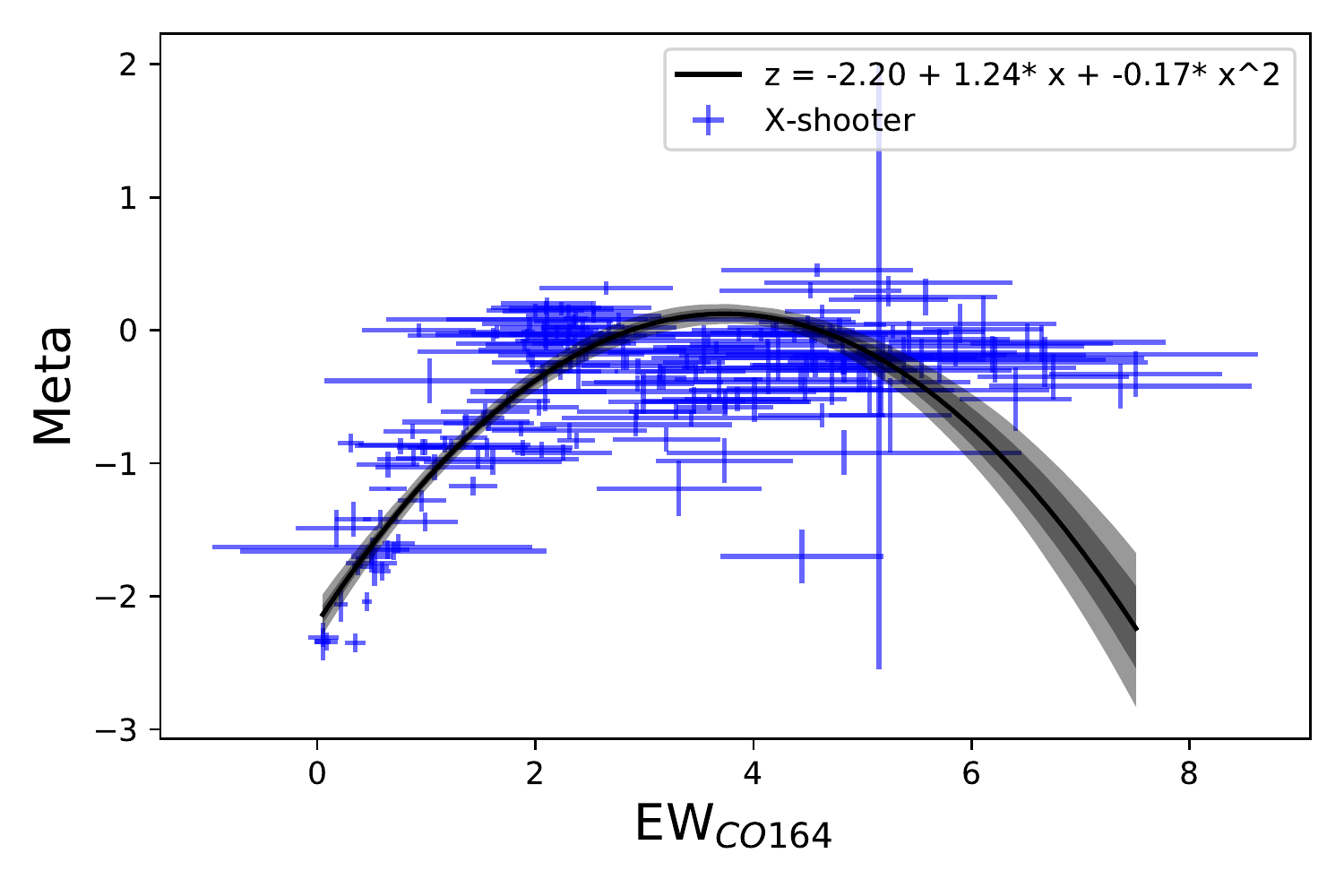} \\
	\includegraphics[scale = 0.15]{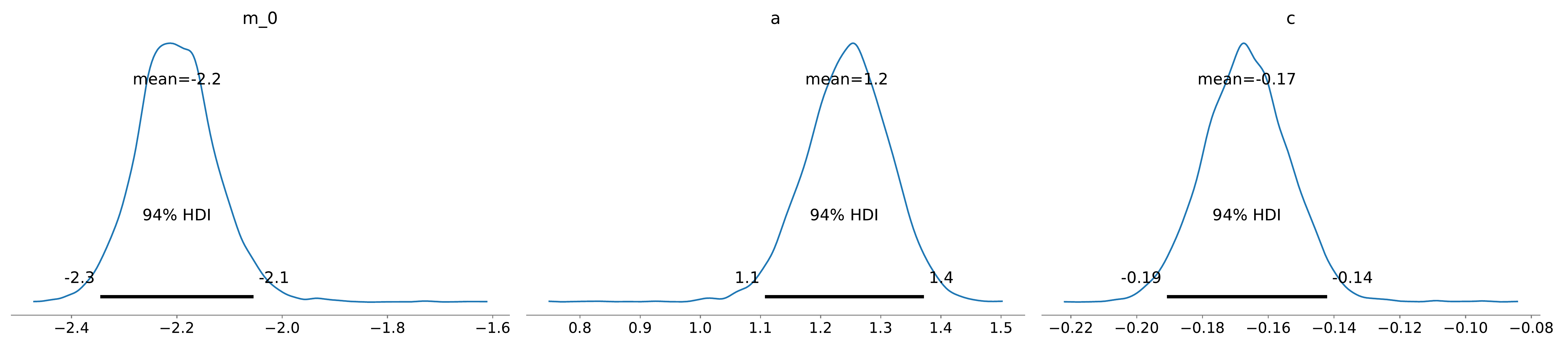} \\			 			
	\includegraphics[scale = 0.45]{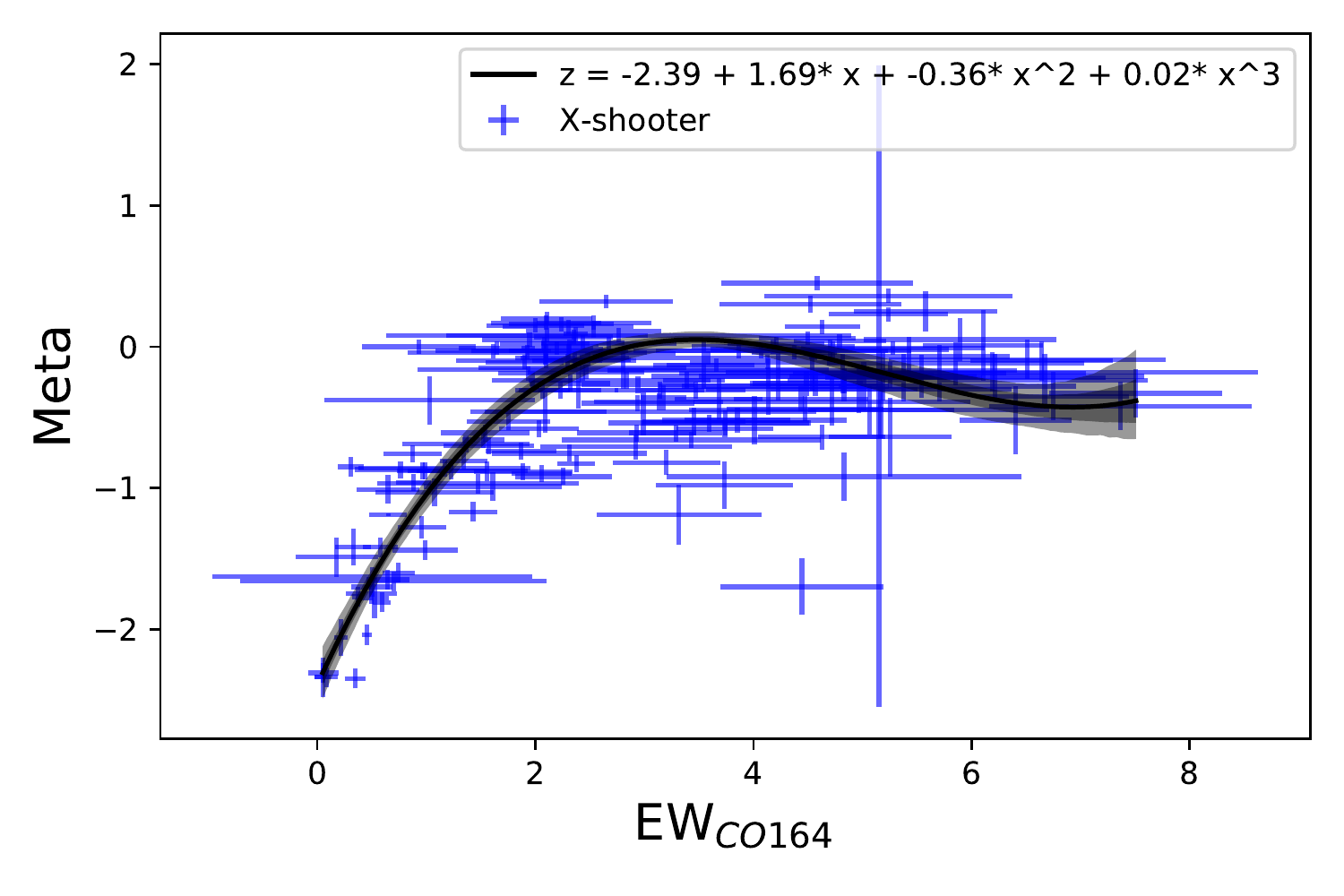} \\
	\includegraphics[scale = 0.12]{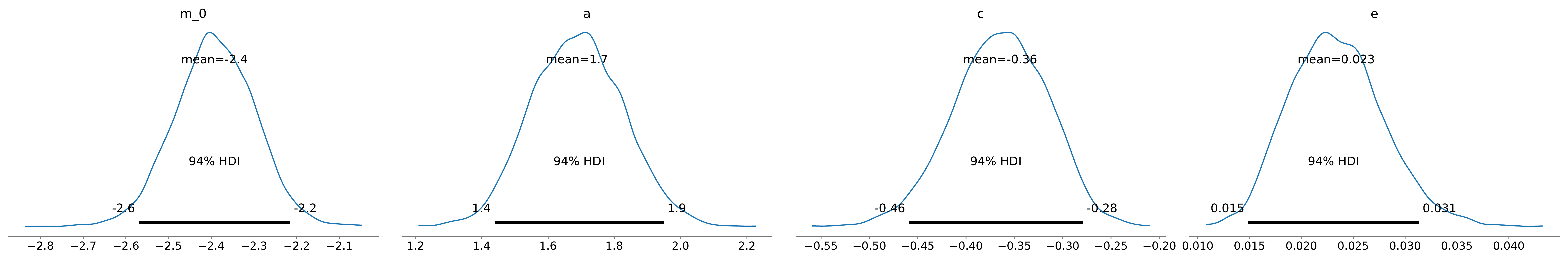} \\ 			
	\caption{Similar to Fig. \ref{Figure:meta_FeH_BayesianFit}, but for CO164.} 
	\label{Figure:meta_CO164_BayesianFit} 
\end{figure}

\begin{figure} 
	\centering
	\includegraphics[scale = 0.45]{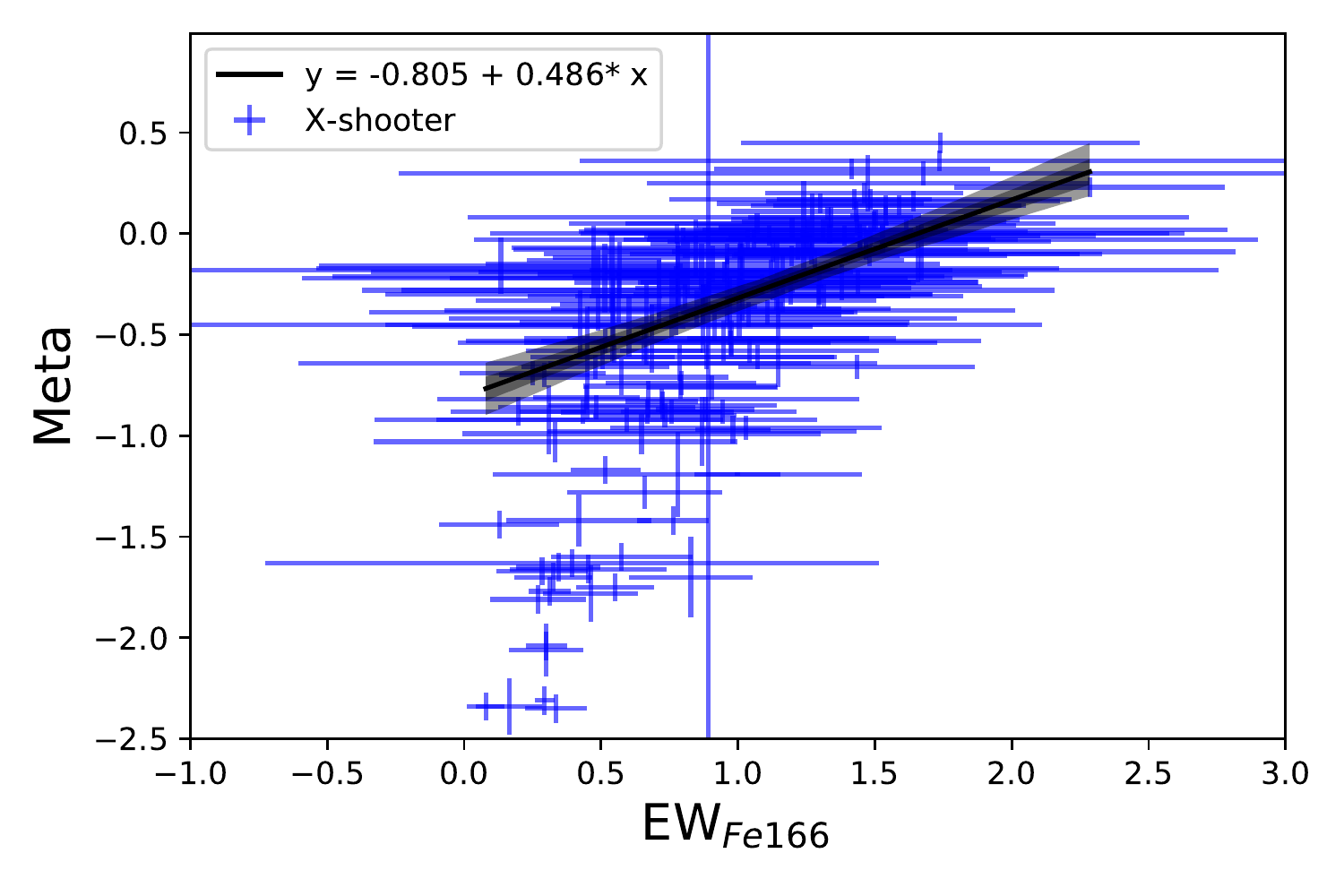} \\
	\includegraphics[scale = 0.16]{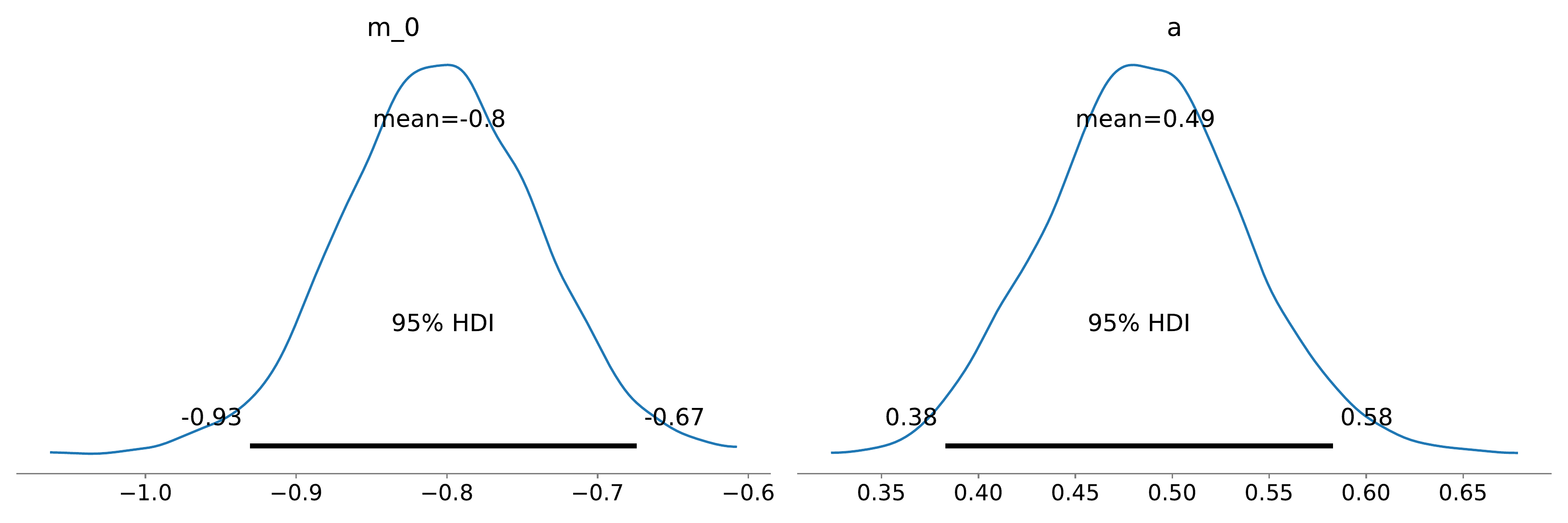} \\			
	\includegraphics[scale = 0.45]{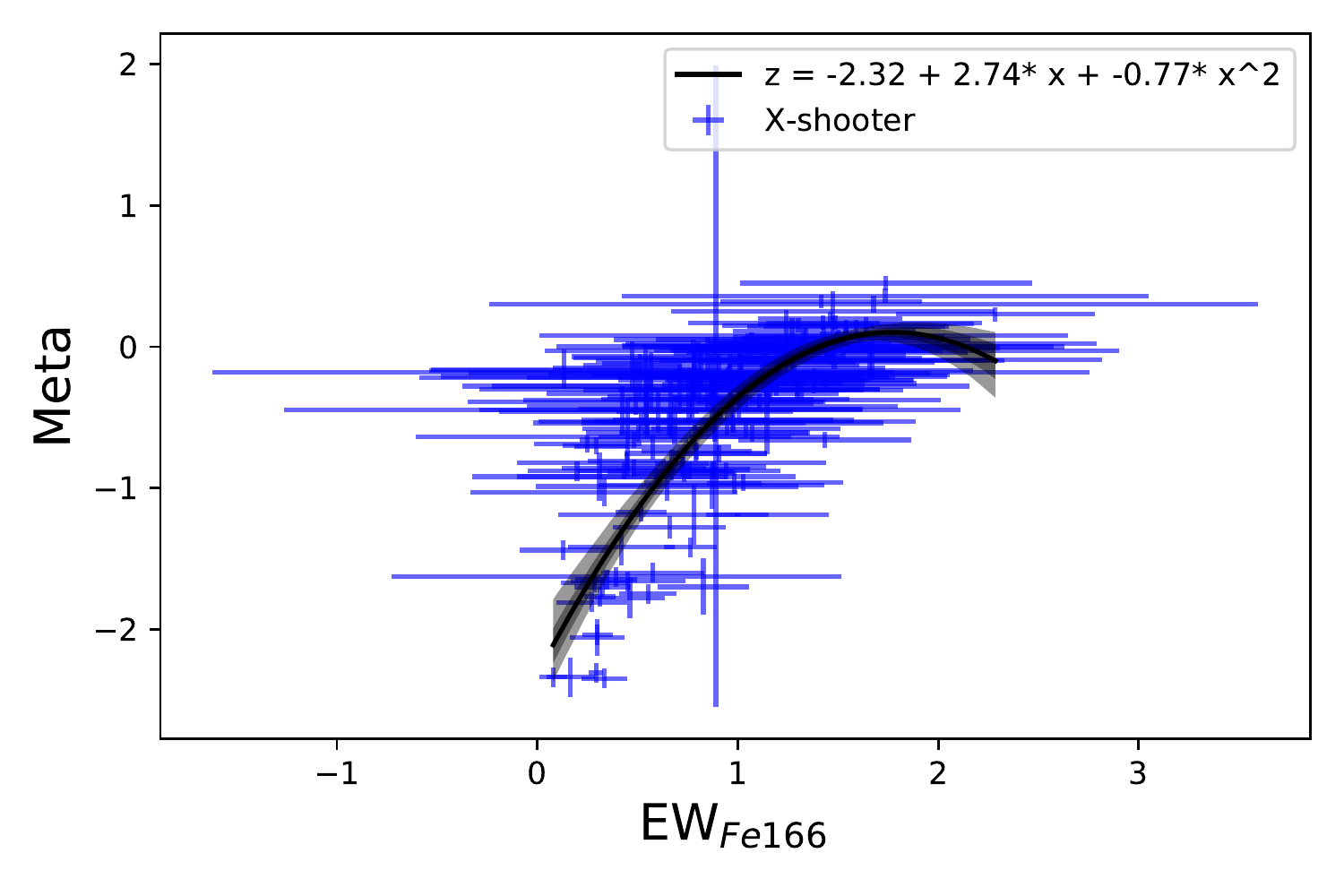} \\
	\includegraphics[scale = 0.15]{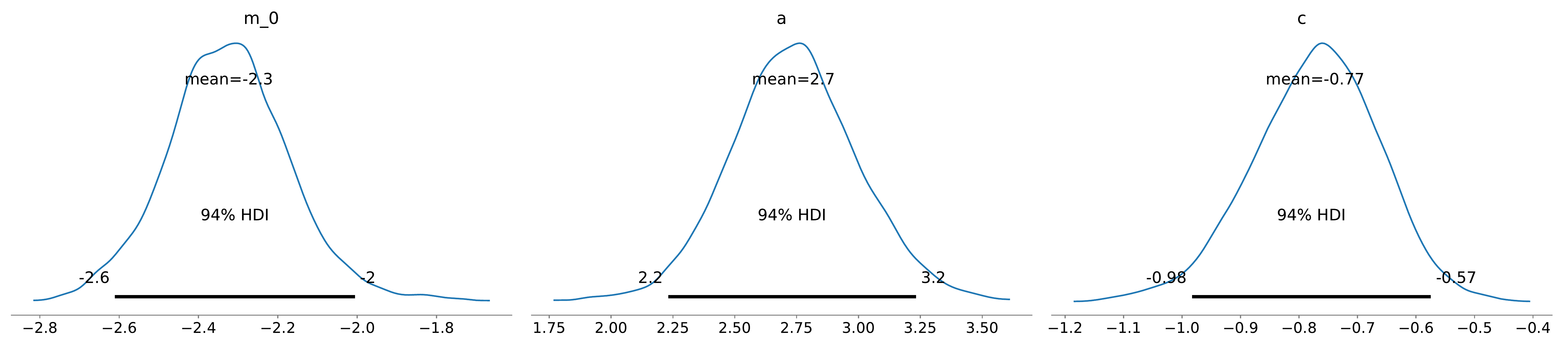} \\			 			
	\includegraphics[scale = 0.45]{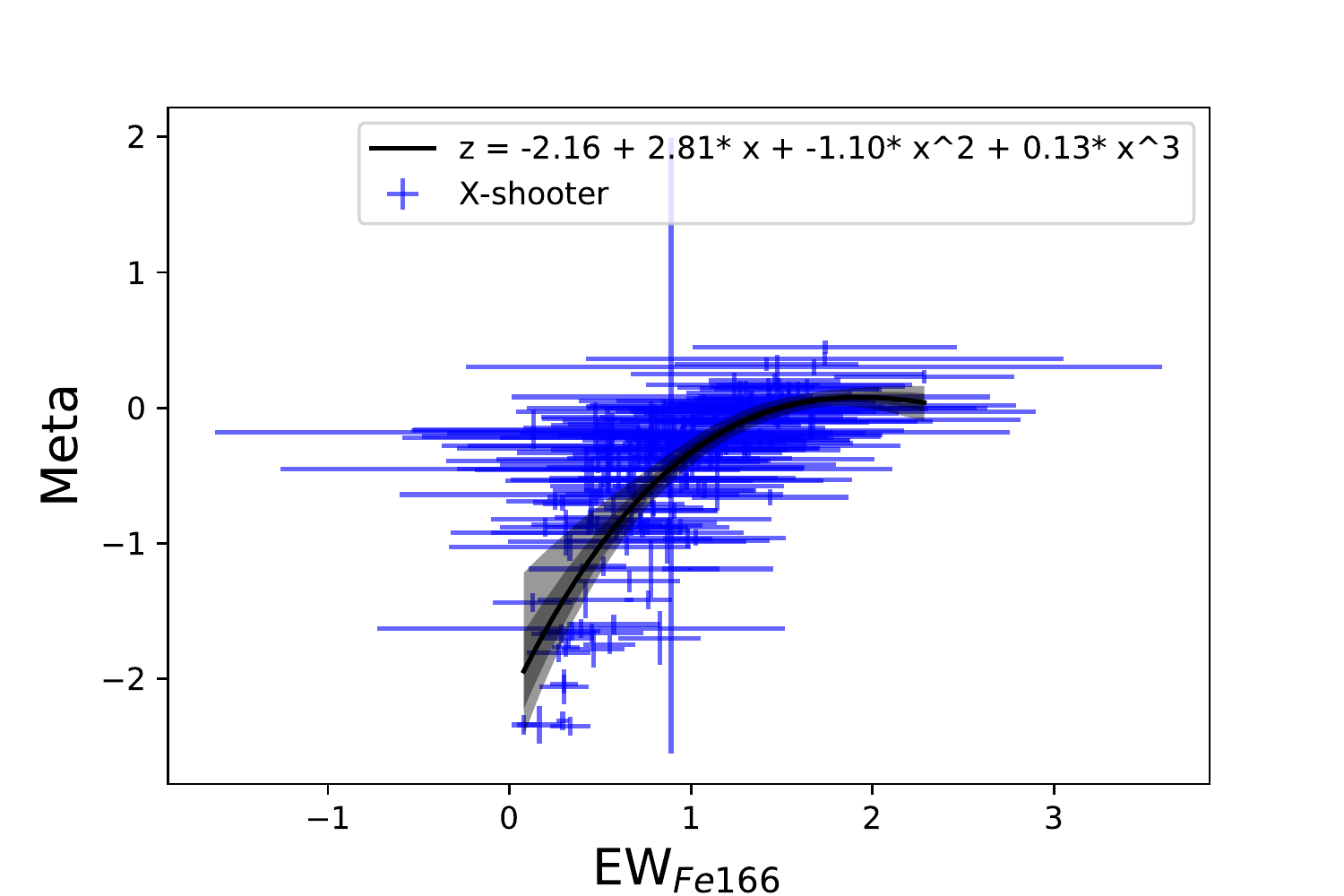} \\
	\includegraphics[scale = 0.12]{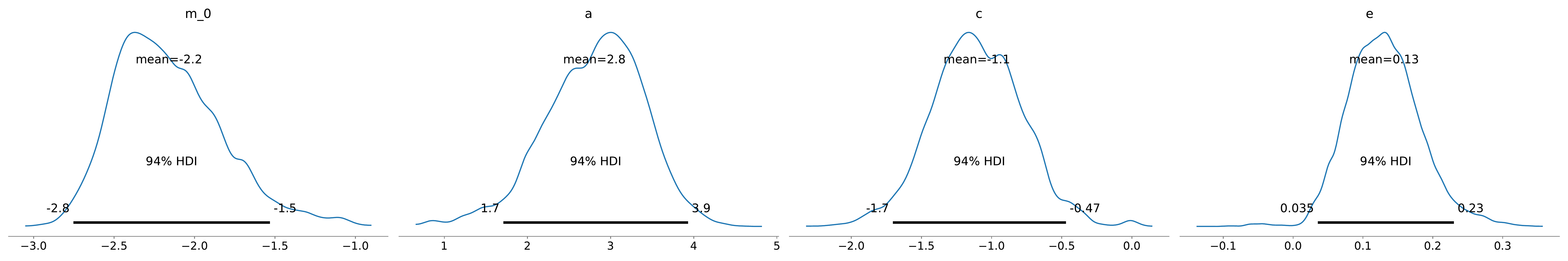} \\ 			
	\caption{Similar to Fig. \ref{Figure:meta_FeH_BayesianFit}, but for Fe166.} 
	\label{Figure:meta_Fe166_BayesianFit} 
\end{figure}

\begin{figure} 
	\centering
	\includegraphics[scale = 0.4]{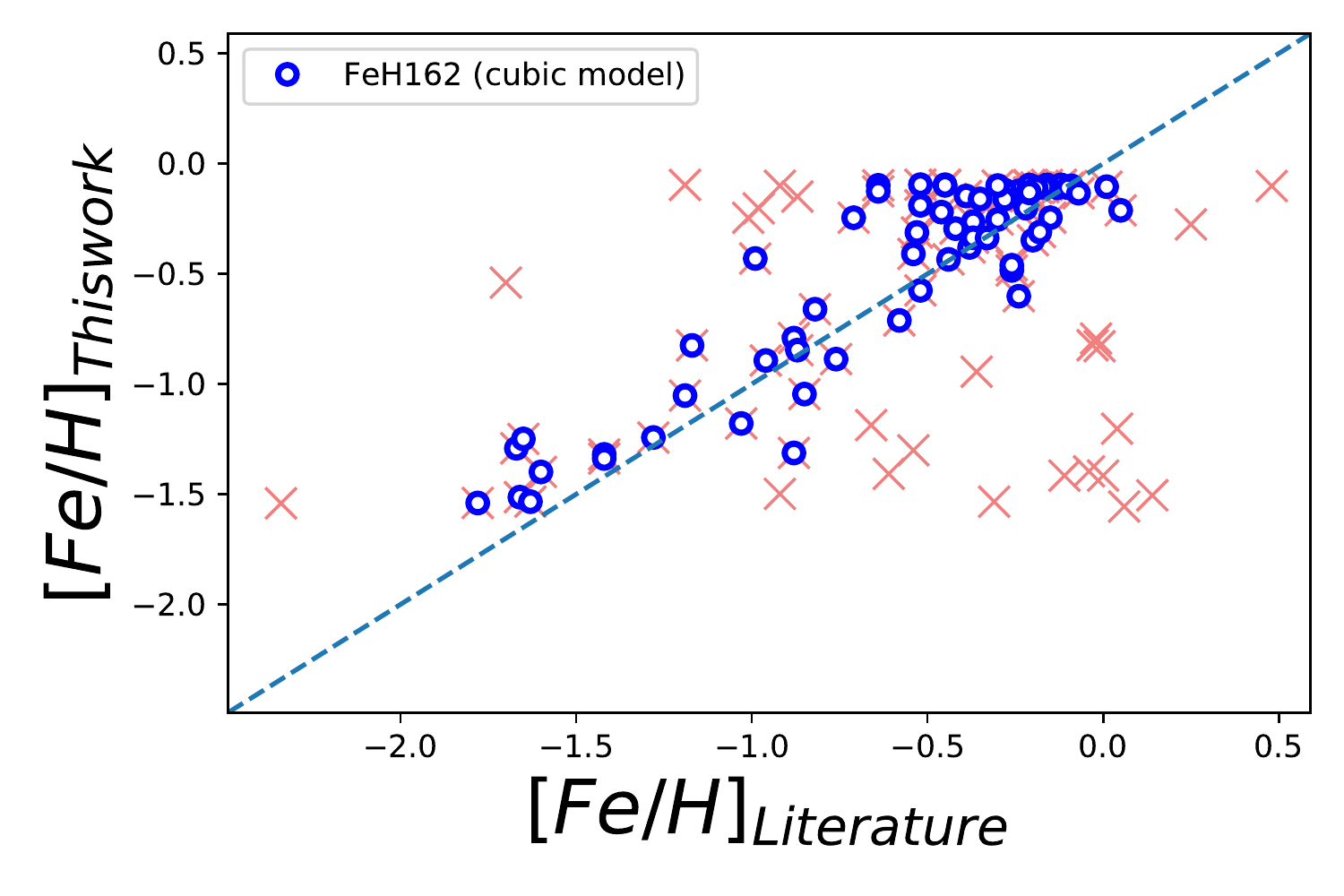} \\
	\includegraphics[scale = 0.4]{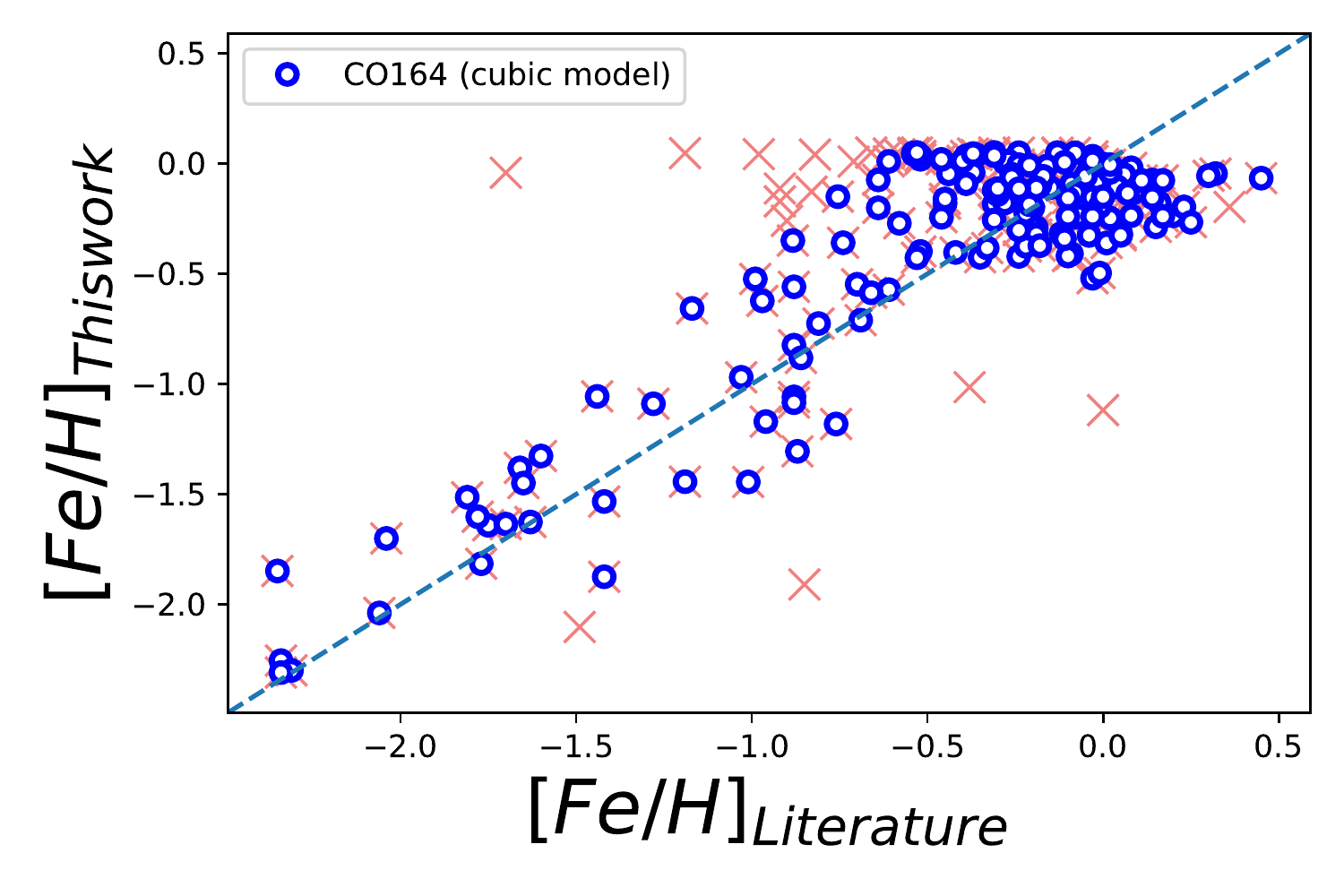} \\
	\includegraphics[scale = 0.4]{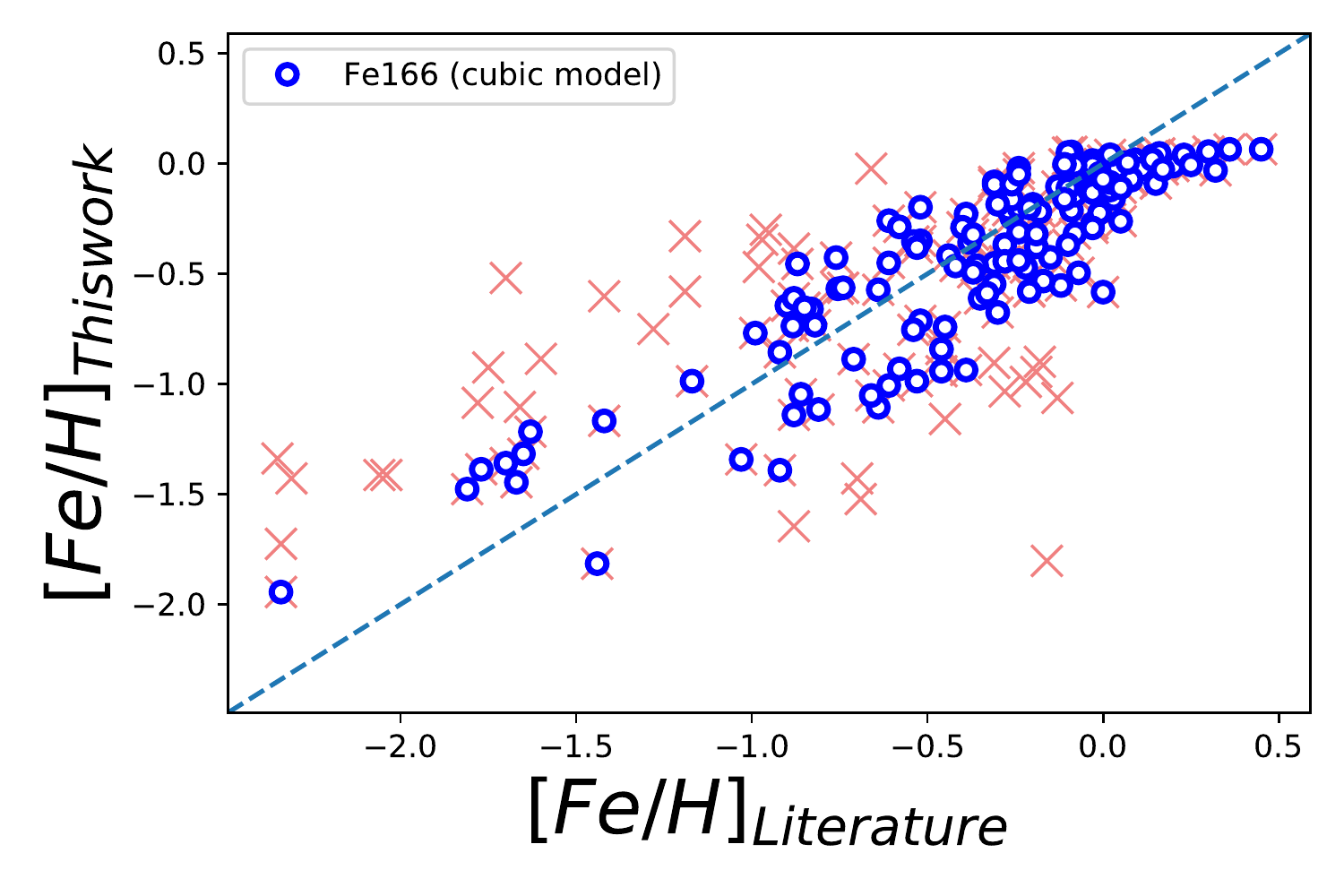}	
	\caption{[$Fe/H$] comparison between derived values from the empirical relations and literature values (upper panel: FeH162, middle panel: CO164, lower panel: Fe166). The coefficients of empirical relations are listed in Table \ref{tab:BayesianModelForMetallicity}. The dashed line displays the one-to-one correspondence of [$Fe/H$]. The red cross symbols represent all of our samples and the blue circles show stars after 2$\sigma$ clipping.} 
	\label{Figure:meta_CO164-Fe166_BayesianFit_residual} 
\end{figure}

 This study confirms the significant influence of metallicity on $T_{eff}$--EWs empirical relations for subsolar stars. However, such influence is not apparent for supersolar stars. This may be because of the narrow metallicity range of the supersolar group and a very small number (4) of stars. A further investigation is recommended for a sufficiently large number of metal-rich stars. Thus, we do not discuss the supersolar group further. 
 
 Furthermore, we estimated $T_{eff}$ to each star (excluding outliers) in each metallicity group using the corresponding empirical solutions, and compare it with literature $T_{eff}$ as shown in Fig. \ref{Figure:1to1_correspondence_for_Hierarchical Bayesian_model}. The mean and standard deviation of the fit residuals are  (i) $\Delta$$T_{eff,Avg}$ = $-$14 K, $\sigma_{T_{eff}}$ = 116 K and $\Delta$$T_{eff,Avg}$ = $-$15 K, $\sigma_{T_{eff}}$ = 92 K for $T_{eff}$--CO156, (ii) $\Delta$$T_{eff,Avg}$ = $-$1 K, $\sigma_{T_{eff}}$ = 56 K and $\Delta$$T_{eff,Avg}$ = $-$12 K, $\sigma_{T_{eff}}$ = 116 K for $T_{eff}$--CO162, and (iii) $\Delta$$T_{eff,Avg}$ = $-$22 K, $\sigma_{T_{eff}}$ = 84 K and $\Delta$$T_{eff,Avg}$ = $-$15 K, $\sigma_{T_{eff}}$ = 94 K for $T_{eff}$--COMg171, for solar and subsolar metallicity group, respectively.
 
 We found from this study that the best accuracy (SEE$_{2\sigma clip}$ = 56 K) in an empirical relationship is provided by CO162 for the solar metallicity group stars, and CO156 (SEE$_{2\sigma clip}$ = 94 K) yield more accurate relation for subsolar stars. On the other hand, CO156 show relatively less sensitivity to metallicity than the other two. Moreover, CO156 was found to be one of best three $T_{eff}$ predictors in the entire parameter space (see Fig.~\ref{Fig:best_linearmodel_for_Teff}). Thus, we can implement $T_{eff}$--CO156 empirical relation in general with a typical accuracy of $\sim$ 150 K at $T_{eff}$ $<$ 4700 K. Otherwise, the choice of empirical relation in evaluating $T_{eff}$ can be made depending on the knowledge of the metallicity. 
 
 \subsubsection{Posterior distribution and metallicity indicators}

Three spectral features, FeH162, CO164, and Fe166, were adopted for establishing empirical relations on metallicity. Equations \ref{equation:bayesian_linear_individualline}, \ref{equation:bayesian_quadratic_individualline}, and \ref{equation:bayesian_cubic_individualline} were adopted for linear, quadratic, and cubic models for [$Fe/H$] versus spectral features relationship, respectively. For this analysis, we used the values of informative priors of model parameters as mentioned in Sec \ref{Section:Specifying prior distribution on parameters}. Figs \ref{Figure:meta_FeH_BayesianFit}, \ref{Figure:meta_CO164_BayesianFit} and \ref{Figure:meta_Fe166_BayesianFit} display Bayesian fits for FeH162, CO164, and Fe166, respectively. Table \ref{tab:BayesianModelForMetallicity} lists the summary of posterior distributions of model coefficients, along with BIC and SEE$_{2\sigma clip}$ of the model fit. The BIC measures the trade-off between model fit and complexity of the model, and the lower BIC value indicates a better fit. We found that the cubic model provides the minimum BIC value in comparison with linear and quadratic models for all cases. $\Delta$BIC$_{linear - cubic}$ (difference in BIC values between linear and cubic models) are 40, 141, and 59, and 
and  $\Delta$BIC$_{quadratic - cubic}$ (difference in BIC values between quadratic and cubic models) are 10, 48, and 22 for FeH162, CO164, and Fe166, respectively. As the differences are significant for the cubic model, we can conclude that the cubic model yields the best empirical relation in case of each spectral feature. The typical accuracy of the best metallicity model (SEE$_{2\sigma clip}$) is 0.22 dex for FeH162, 0.28 dex for CO164, and 0.24 dex Fe166.

We then invert the process and calculate [$Fe/H$] of the sample giants using the above established cubic models. The comparison of our measurements with the literature value is illustrated in  Fig. \ref{Figure:meta_CO164-Fe166_BayesianFit_residual}, where the `X' symbols and blue circles represent the giants before and after 2$\sigma$ clipping, respectively. The mean and standard deviation of the fit residuals are (i) $\Delta$[$Fe/H$]$_{Avg}$ = $-$0.09 dex and $\sigma_{[Fe/H]}$ = 0.20 dex for FeH162, (ii) $\Delta$[$Fe/H$]$_{Avg}$ = $-$0.03 dex and $\sigma_{[Fe/H]}$ = 0.28 dex for CO164, and (iii) $\Delta$[$Fe/H$]$_{Avg}$ = 0.07 dex and $\sigma_{[Fe/H]}$ = 0.23 dex for Fe166. 

\begin{table*}
	\centering
	\caption{Posteriors distribution of Bayesian models for [$Fe/H$] versus EWs of spectral features relationship and  comparison between various [$Fe/H$] correlations.}
	\label{tab:BayesianModelForMetallicity}
	\resizebox{0.99\textwidth}{!}{
		\begin{tabular}{lcccccccccccclc} 
			\hline
			\hline
			Index & T & N & m0$^a$ & a$^a$ & b$^a$ & c$^a$ & d$^a$ & e$^a$ & BIC & SEE & N$_{2\sigma clip}$ & SEE$_{2\sigma clip}$ & Relation$^a$ & Range$_{2\sigma clip}$$^b$ \\	
			\hline

			x = EW$_{FeH162}$ & 177 & 93 & $-$0.889 $\pm$ 0.082 & 0.259 $\pm$ 0.042 & ... & ... & ... & ... & 204 & 0.48 & 83 & 0.39 & Equation (\ref{equation:bayesian_linear_individualline}) & [$-$1.66, 0.48] \\
			x = EW$_{FeH162}$ & 177 & 93 & $-$1.434 $\pm$ 0.087 & 1.179 $\pm$ 0.132 & ... & $-$0.262 $\pm$ 0.042 & ... & ... & 174 & 0.53 & 73 & 0.27 & Equation (\ref{equation:bayesian_quadratic_individualline}) & [$-$1.78, 0.48] \\
			x = EW$_{FeH162}$ & 177 & 93 & $-$1.563 $\pm$ 0.136 & 1.471 $\pm$ 0.293 & ...  & $-$0.460 $\pm$ 0.153 & ... & 0.042 $\pm$ 0.021 & 164 & 0.55 & 69 & 0.22 & Equation (\ref{equation:bayesian_cubic_individualline}) & [$-$1.78, 0.05]  \\
			&&&&&&&&&&&&&& \\
			
			x = EW$_{CO164}$ & 177 & 171 & $-$1.078 $\pm$ 0.052 & 0.219 $\pm$ 0.017 & ... & ... & ... & ... & 380 & 0.51 & 162 & 0.44 & Equation (\ref{equation:bayesian_linear_individualline}) & [$-$1.81, 0.45] \\
			x = EW$_{CO164}$ & 177 & 171 & $-$2.199 $\pm$ 0.080 & 1.244 $\pm$ 0.073 & ... & $-$0.166 $\pm$ 0.013 & ... & ... & 287 & 0.50 & 151 & 0.34 & Equation (\ref{equation:bayesian_quadratic_individualline}) & [$-$2.35, 0.45] \\
			x = EW$_{CO164}$ & 177 & 171 & $-$2.392 $\pm$ 0.095 & 1.688 $\pm$ 0.138 & ...  & $-$0.365 $\pm$ 0.049 & ... & 0.023 $\pm$ 0.005 & 239 & 0.39 & 154 & 0.28 & Equation (\ref{equation:bayesian_cubic_individualline}) & [$-$2.35, 0.45]  \\ 
			&&&&&&&&&&&&&& \\
			x = EW$_{Fe166}$ & 177 & 170 & $-$0.805 $\pm$ 0.065 & 0.486 $\pm$ 0.051 & ... & ... & ... & ... & 331 & 0.50 & 131 & 0.20 & Equation (\ref{equation:bayesian_linear_individualline}) & [$-$0.92, 0.36] \\
			x = EW$_{Fe166}$ & 177 & 170 & $-$2.323 $\pm$ 0.159 & 2.737 $\pm$ 0.264 & ... & $-$0.772 $\pm$ 0.109 & ... & ... & 294 & 0.41 & 143 & 0.26 & Equation (\ref{equation:bayesian_quadratic_individualline}) & [$-$2.34, 0.45] \\
			x = EW$_{Fe166}$ & 177 & 170 & $-$2.159 $\pm$ 0.336 & 2.806 $\pm$ 0.595 & ...  & $-$1.105 $\pm$ 0.326 & ... & 0.130 $\pm$ 0.053 & 272 & 0.40 & 140 & 0.24 & Equation (\ref{equation:bayesian_cubic_individualline}) & [$-$2.34, 0.45]  \\
			
			&&&&&&&&&&&&&& \\
			Meta < $-$ 0.3 dex &&&&&&&&&&&&&& \\
			x = EW$_{CO164}$ & 177 & 86 & $-$1.046 $\pm$ 0.059 & 0.130 $\pm$ 0.021 & ... & ... & ... & ... & 156 & 0.48 & 69 & 0.24 & Equation (\ref{equation:bayesian_cubic_individualline}) & [$-$1.49, -0.31] \\
			x = EW$_{CO164}$ & 177 & 86 & $-$2.142 $\pm$ 0.108 & 1.074 $\pm$ 0.125 & ... & $-$0.161 $\pm$ 0.029 & ... & ... & 115 & 0.59 & 74 & 0.24 & Equation (\ref{equation:bayesian_cubic_individualline}) & [$-$2.34, $-$0.31] \\
			x = EW$_{CO164}$ & 177 & 86 & $-$2.336 $\pm$ 0.100 & 1.514 $\pm$ 0.155 & ...  & $-$0.372 $\pm$ 0.059 & ... & 0.028 $\pm$ 0.006 & 92 & 31 & 75 & 0.19 & Equation (\ref{equation:bayesian_cubic_individualline}) & [$-$2.34, $-$0.31]  \\
			&&&&&&&&&&&&&& \\	
			
			\hline
	\end{tabular}} \\	
	T $-$ total nos. of data points; N $-$ no. of points having EWs $>$ 0; N $_{2\sigma clip}$ $-$ no. of points after 2$\sigma$ clipping of N \\
	SEE $-$ standard error of estimate before 2$\sigma$ clipping; 
	SEE$_{2\sigma clip}$ $-$ standard error of estimate after 2$\sigma$ clipping.\\ 
	$^a$ Relation (equation) used to establish the correlation; m0, a, b, c, d, e are coefficients of the equation.\\ $^b$ [$Fe/H$] range of the stars after 2$\sigma$ clipping.
	\\
\end{table*}

Comparing SEE$_{2\sigma clip}$ values, it appears that the best empirical relation for estimating metallicity is given by FeH162. However, from Fig. \ref{Figure:meta_CO164-Fe166_BayesianFit_residual}, we can see that the measured metallicity of many stars is either overestimated or underestimated in case of FeH162 and Fe166. The reason could be manifold. For Fe166, most of them are metal-poor ($[$Fe/H$]$ $<$ $-$1.0 dex). In addition, some stars having [$Fe/H$] $<$ $-$1.0 dex are relatively warmer giants ($T_{eff}$ $>$ 4500 K). Thus, it appears that Fe166 is mainly sensitive to metallicity at [$Fe/H$] $<$ $-$1.0 dex with $T_{eff}$ $<$ 4500 K. For FeH162, the discrepancy could be because of the contamination by strong OH features at 1.625 $\mu$m as most of them have either metallicity below $-$0.5 dex and/or effective temperature below $<$ 4000 K. On the other hand, predicted metallicity from CO164 is in good agreement with literature below $-$0.3 dex. As can be seen in Fig. \ref{Figure:meta_CO164_BayesianFit}, the sensitivity of CO164 to [Fe/H] appears to decrease at [$Fe/H$] $>$ $-$0.3 dex. This may be because of the saturation of CO164 for solar-metallicity giants. Thus, we narrow down the metallicity range by considering only those stars having [$Fe/H$] $\leq$ $-$0.3 dex. Three Bayesian models (linear, quadratic, and cubic) were also exercised here. The summary of the posterior distribution of three models is tabulated in Table \ref{tab:BayesianModelForMetallicity} along with estimated BIC and SEE values. The cubic model yielded the best relationship with a typical accuracy of 0.19 dex. Thus, the accuracy of the metallicity empirical relation improves significantly by narrowing down the metallicity range. 

In addition, we explored different combinations of spectral lines to improve the accuracy of metallicity empirical relations using the equations \ref{equation:bayesian_linear_combinedlines} and \ref{equation:bayesian_quadratic_combinedlines}. However, no multi-lines solution yields a significant improvement in the accuracy of the empirical relationship as discussed above.

\section{Summary and Conclusions} \label{Summary_and_Conclusions}

In this paper, we explored the $H$-band atmospheric window aiming to provide easy-to-use quantitative diagnostic tools for estimating stellar parameters ($T_{eff}$ and [$Fe/H$]) of cool giants from low-resolution spectra. For that, we obtained a total of 177 cool giants ($<$ 5000 K) of $R$ $\sim$ 10000 from the X-shooter spectral library having a wider metallicity coverage (from $-$2.35 dex to 0.5 dex) than in earlier works. After degrading the spectral resolution to R $\sim$ 1200, we measured equivalent widths of 20 spectral features, for instance, Si I at 1.59 $\mu$m, $^{12}$CO at 1.62 and 1.64 $\mu$m, Fe I at 1.66 $\mu$m, Al I at 1.67 $\mu$m, and COMg line at 1.71 $\mu$m. We first investigated the behavior of spectral features with $T_{eff}$ and [$Fe/H$] and then derived empirical relationships between equivalent widths and stellar parameters in the Bayesian framework. The main results in this work are summarized as follows.

\begin{enumerate}
\item We found that the most of the spectral features studied here show strong dependence on $T_{eff}$. They show negative correlation with $T_{eff}$ except H {\small I} at 1.68 $\mu$m. K {\small I} at 1.52 $\mu$m, H {\small I} 1.55 $\mu$m, $^{12}$CO at 1.56 $\mu$m, $^{12}$CO at 1.62 $\mu$m, FeH at 1.62 $\mu$m, $^{12}$CO at 1.66 $\mu$m, Al {\small I} at 1.67 $\mu$m, H {\small I} at 1.68 $\mu$m and COMg at 1.71 $\mu$m appear as good indicators for $T_{eff}$.

\item FeH at 1.62 $\mu$m, $^{12}$CO at 1.64 $\mu$m, and Fe {\small I} at 1.66 $\mu$m are strongly sensitive to [$Fe/H$] and can be used as metallicity predictors. However, many other metallic lines, for example Mg {\small I} at 1.50 $\mu$m, K {\small I} at 1.52 $\mu$m, Mg {\small I} at 1.57 $\mu$m, Fe {\small I} at 1.61 $\mu$m, and Al {\small I} at 1.67 $\mu$m, show very weak or no correlation with metallicity.

\item We established new empirical relations between effective temperature and equivalent widths of several spectral features. Among them, $^{12}$CO at 1.56 $\mu$m and 1.62 $\mu$m, and CO+Mg {\small I} at 1.71 $\mu$m are the best three $T_{eff}$ indicators with a typical accuracy of 153 K, 123 K, and 107 K, respectively. In addition, we showed a detailed quantitative metallicity dependence of these empirical relations defining three metallicity groups, supersolar ([$Fe/H$] > 0.0 dex), solar ($-$0.3 dex $<$ [$Fe/H$] $<$ 0.3 dex), and subsolar ([$Fe/H$] $<$ $-$0.3 dex), from Hierarchical Bayesian modelling. We found that the difference between solar and subsolar $T_{eff}$--EWs empirical relationship is statistically significant. Thus, we can conclude that the derived $T_{eff}$ for metal-poor stars using the empirical relations based on solar-neighborhood stars is not completely reliable. However, the supersolar group shows no statistically significant difference from the solar group.  Furthermore, $^{12}$CO at 1.56 $\mu$m shows relatively less sensitivity on the metallicity than$^{12}$CO at 1.62 $\mu$m and COMg {\small I} at 1.71 $\mu$m, and thus, it could be used more generally.

\item We adopted FeH at 1.62 $\mu$m, $^{12}$CO at 1.64 $\mu$m, and Fe {\small I} at 1.66 $\mu$m for obtaining new empirical relations between metallicity and the spectral features. We found that the cubic Bayesian model provides the best metallicity estimator with a typical accuracy of 0.22 dex for FeH at 1.62 $\mu$m, 0.28 dex for $^{12}$CO at 1.64 $\mu$m and 0.24 dex for Fe {\small I} at 1.66 $\mu$m. To our knowledge, this is the first metallicity scale for cool giants in the $H$-band atmospheric window.

\end{enumerate}  
We expect that this work will help to understand the behaviour of spectral features in a relatively less explored $H$-band atmospheric window with stellar parameters in a greater detail. Our provided diagnostic tools can be implemented very easily to derive stellar parameters, specifically effective temperature and metallicity of cool giants from the low-resolution spectra without any knowledge of the reddening and distance to the object. In addition, such diagnostic tools are very helpful in survey mode observations with readily available spectra.

\section*{Acknowledgements}
The authors are thankful to the reviewer for their critical and valuable comments, which helped us to improve the paper. This research work is supported by the Tata Institute of Fundamental Research, Mumbai under the Department of Atomic Energy, Government of India. SG and DKO acknowledge the support of the Department of Atomic Energy, Government of India, under project Identification No. RTI 4002. This research has made use of the SIMBAD database, operated at CDS, Strasbourg, France.

\section*{DATA AVAILABILITY}
All observational data utilized in this paper are publicly available and can be found at: \url{http://xsl.astro.unistra.fr/} (X-shooter data). Table~\ref{tab:Stellar_parameters_and_estimated_EWs} is available in its entirety as online supplementary material.




\bibliographystyle{mnras}
\bibliography{supriyo_reference-v1.bib}







\bsp	
\label{lastpage}

\end{document}